\documentclass[
reprint,
superscriptaddress,
preprintnumbers,
 amsmath,
 amssymb,
 prC,
]{revtex4-1}

\usepackage[     bookmarks,
                 bookmarksopen = true,
                 bookmarksnumbered = true,
                 linktocpage,
                 colorlinks = true,
                 linkcolor = blue,
                 urlcolor  = blue,
                 citecolor = blue,
                 anchorcolor = green,
                 hyperindex = true,
                 hyperfigures]
        {hyperref}

\usepackage{graphicx}
\usepackage{dcolumn}
\usepackage{bm}
\usepackage{float}
\usepackage{amsmath}
\usepackage{soul}

\usepackage{color}
\usepackage{graphicx}
\definecolor{dgreen}{cmyk}{1.,0.,1.,0.2}        
\definecolor{orange}{cmyk}{0.,0.353,1.,0.}    

\def \c {\chi}

\begin{document}

\title{Nuclear equation of state at finite $\mu_B$ using deep learning assisted quasi-parton model}






\author{Fu-Peng Li}
\affiliation{Key Laboratory of Quark and Lepton Physics (MOE) \& Institute of Particle Physics, Central China Normal University, Wuhan 430079, China}

\author{Long-Gang Pang}
\email[]{lgpang@ccnu.edu.cn}
\affiliation{Key Laboratory of Quark and Lepton Physics (MOE) \& Institute of Particle Physics, Central China Normal University, Wuhan 430079, China}

\author{Guang-You Qin}
\email[]{guangyou.qin@ccnu.edu.cn}
\affiliation{Key Laboratory of Quark and Lepton Physics (MOE) \& Institute of Particle Physics, Central China Normal University, Wuhan 430079, China}


\date{\today}%

\begin{abstract}

To accurately determine the nuclear equation of state (EoS) at finite baryon chemical potential ($\mu_B$) remains a challenging yet essential goal in the study of QCD matter under extreme conditions. In this study, we develop a deep learning assisted quasi-parton model, which utilizes three deep neural networks, to reconstruct the QCD EoS at zero $\mu_B$ and predict the EoS and transport coefficient $\eta/s$ at finite $\mu_B$. The EoS derived from our quasi-parton model shows excellent agreement with lattice QCD results obtained using Taylor expansion techniques. The minimum value of $\eta/s$ is found to be approximately 175 MeV and decreases with increasing chemical potential within the confidence interval. This model not only provides a robust framework for understanding the properties of the QCD EoS at finite $\mu_B$ but also offers critical input for relativistic hydrodynamic simulations of nuclear matter produced in heavy-ion collisions by the RHIC beam energy scan program.

\end{abstract}
\maketitle

\section{Introduction}
Reconstructing the nuclear equation of state (EoS) is a cornerstone of modern nuclear and particle physics, with profound implications for understanding the evolution of the early universe, the properties of neutron stars, and the behavior of Quark-Gluon Plasma (QGP) \cite{Akmal:1998cf,Lattimer:2000nx,Lattimer:2012nd,Braun-Munzinger:2015hba,ALICE:2016fzo,Burgio:2021vgk}. This pursuit is a central focus of heavy-ion collision experiments \cite{Sorensen:2023zkk,Arslandok:2023utm,MUSES:2023hyz}, which have driven significant advancements in our understanding of the EoS of hot and dense nuclear matter over the past few decades \cite{Bazavov:2011nk,Fukushima:2010bq,Baym:2017whm,Andersen:2014xxa,Fukushima:2013rx}. Notably, first-principles lattice QCD calculations at zero baryon chemical potential ($\mu_B$) have provided a critical benchmark for validating theoretical models \cite{HotQCD:2014kol,HotQCD:2018pds,Bazavov:2017dus,HotQCD:2012fhj,Bazavov:2020bjn}.

Despite these achievements, the exploration of QCD at finite baryon chemical potentials remains a formidable challenge, primarily due to the notorious sign problem \cite{deForcrand:1999fz,Engels:1999id,Cox:1999nt}. While effective expansion methods offer interpolation and extrapolation capabilities \cite{Bollweg:2022rps,Borsanyi:2021sxv,Monnai:2019hkn}, they introduce significant uncertainties, particularly at higher baryon chemical potentials. Alternative approaches, such as the hadron resonance gas model and quasi-parton models, have been employed to approximate the QCD EoS \cite{Vovchenko:2016rkn,Vovchenko:2019pjl,Monnai:2024pvy}. However, achieving a precise and comprehensive understanding of EoS across all relevant conditions remains an unresolved and critical problem in the field.

The recent surge in machine learning (ML) applications has revolutionized scientific computation, particularly in solving inverse problems in physics through data-driven and physics-driven neural networks \cite{2018PINN, pde712178, khoo_lu_ying_2021, pde1528518, 2017DGM, 2018The, 2021Deep, RAISSI2019686, Soma:2022vbb,Zhou:2023pti,Pang:2016vdc,Boehnlein:2021eym,Aarts2025,Pang:2024kid}. Data-driven approaches leverage deep neural networks (DNNs) to extract physical information from data, though they often struggle with extrapolation. In contrast, physics-driven methods integrate the underlying physics theory, such as partial differential equations (PDEs), into the training process, enhancing predictive accuracy. The universal approximation capabilities of neural networks, as demonstrated by Hornik et al. \cite{Hornik1989MultilayerFN}, combined with the precision of automatic differentiation techniques \cite{JMLR:v18:17-468}, have enabled these models to approximate solutions even with limited data. By embedding boundary and initial conditions into the training algorithm, these networks respect the underlying physical constrains, further improving their predictive power.

In this study, we propose a novel weakly coupled quasi-parton model that treats the strongly-interacting QGP as a system of non-interacting quasi-partons. This model derives the QCD EoS using statistical formulae, serving as an effective theory for strongly coupled QGP. By assuming that the masses of quasi-partons (quasi-gluons and quasi-quarks) depend on the temperature and chemical potential of the medium \cite{1976A}, we compute various QCD observables. Recent studies have successfully applied quasi-parton methodologies to heavy-ion collision analyses \cite{Liu:2021dpm,Soloveva:2021quj,Soloveva:2023tvj}. However, these approaches often rely on parameterized mass functions, which can introduce biases. To address this limitation, we employ three neural networks to model the unknown quasi-parton mass functions, determining the masses by minimizing an objective function, as demonstrated in our prior work \cite{Li:2022ozl}. In Ref.~\cite{Li:2022ozl}, the temperature-dependent masses derived from the neural network accurately reproduced the QCD EoS at zero chemical potential.

The paper is organized as follows. Section 2 introduces the quasi-parton method and the neural network framework. Section 3 presents the results and discussion, including training outcomes and predictions at finite $\mu_B$. Finally, Section 4 provides a summary of our findings.

\begin{figure*}[htb]
\begin{centering}
\includegraphics[width=0.8\textwidth]{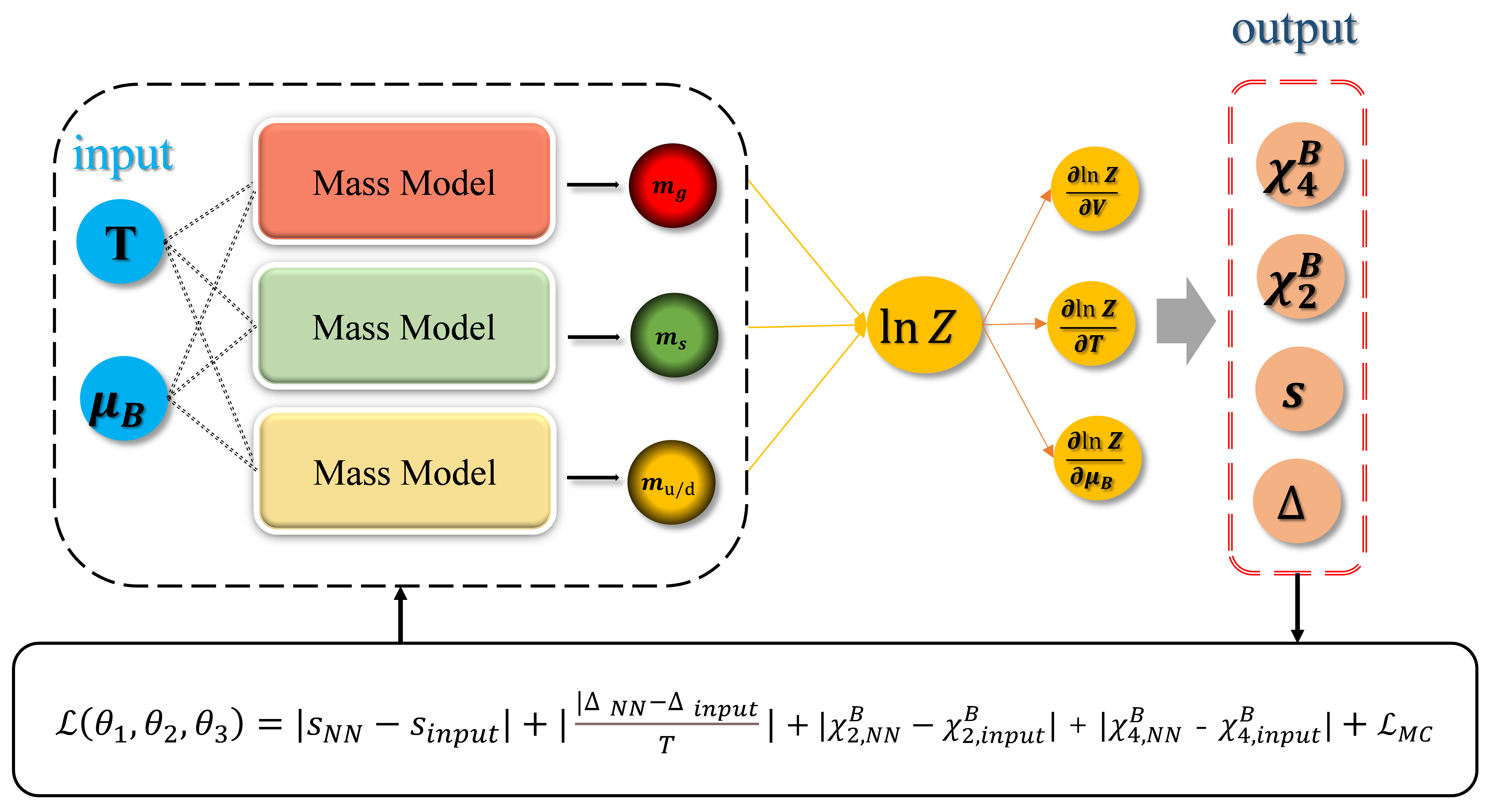}
\caption{(Color online) The neural network framework utilized in this study for deriving three mass functions is structured as follows. Each mass model integrates two residual neural networks (ResNets). Each ResNet consists of 8 hidden layers, with each layer containing 32 neurons. A swish activation function is applied at the conclusion of each hidden layer.}
\label{fig:network}
\end{centering}
\end{figure*}

\section{Method} 
Fig.~\ref{fig:network} shows the temperature ($T$) and chemical potential ($\mu_B$) dependent quasi-parton model developed using PyTorch\cite{NEURIPS2019_bdbca288}. The input data are $T$ and $\mu_B$, and the outputs are $m_{u/d}(T,\mu_B, \theta_1)$, $m_{s}(T,\mu_B, \theta_2)$, and $m_{g}(T,\mu_B, \theta_3)$, respectively. The $\theta_1$, $\theta_2$ and $\theta_3$ represent the trainable parameters associated with these three mass models. For simplicity, we will omit $\theta_i$ and denote $m(T,\mu_B,\theta_i)$ as $m(T, \mu_B)$. Each mass model employs a residual neural network architecture comprising 8 hidden layers, each containing 32 neurons, and utilizes a sigmoid activation function at the output layer to ensure that the results fall within an appropriate range. Once the temperature and baryon chemical potential dependent quasi-parton masses $m_{u/d}(T,\mu_B)$, $m_s(T,\mu_B)$, and $m_g(T,\mu_B)$ are obtained, we can calculate the total partition function $\ln Z(T, \mu_B)$\cite{Kapusta:2006pm}:
\begin{equation}
\ln Z(T,\mu_B) = \ln Z_g(T,\mu_B) + \sum_i \ln Z_{q(\bar{q})_i}(T,\mu_B),
\label{eq:lnz}
\end{equation}
where $i$ represents the quark flavor. We assume that the masses of the quarks and anti-quarks masses are the same. The detailed partition function $\ln Z(T,\mu_B)$ is given by the ideal gas statistical formulae:
\begin{widetext}
\begin{align}
\ln Z_g(T,\mu_B) &= - \frac{d_g V}{2 \pi^{2}} \int_{0}^{\infty}p^{2} dp  \ln \left[ 1 - \exp \left(-{1 \over T}\sqrt{p^{2}+m_g^{2}(T,\mu_B)}\right) \right], \\
\ln Z_{q(\bar{q})_i}(T,\mu_B) &= + \frac{d_{q(\bar{q})_i} V}{2 \pi^{2}} \int_{0}^{\infty}p^{2} dp \ln \left[ 1 + \exp \left(- {1 \over T}\left(\sqrt{p^{2}+m_{q(\bar{q})_i}^{2}(T,\mu_B)} - \mu_{q(\bar{q})_i}\right)\right) \right],
\label{eq:lnz_gluon}
\end{align}
\end{widetext}
The $d_g$ and $d_{q(\bar{q})_i}$ represent the degrees of freedom for the gluon and for the quark and anti-quark, respectively. For quarks and anti-quarks, it is $6 = 3 \times 2 = N_s \times N_c$, where $N_s = 2$ accounts for spin degeneracy and $N_c = 3$ for the number of colors. For gluons, it is $16 = 2 \times 8$. The quark and baryon chemical potentials are related as follows\cite{HotQCD:2012fhj}:
\begin{align}
\mu_u &= \frac{1}{3}\mu_B + \frac{2}{3}\mu_Q,\\
\mu_d &= \frac{1}{3}\mu_B - \frac{1}{3}\mu_Q,\\
\mu_s &= \frac{1}{3}\mu_B - \frac{1}{3}\mu_Q - \mu_S,
\label{eq:muqg}
\end{align}
where in a system with zero net strangeness and zero net electric charge, $\mu_Q$ and $\mu_S$ are zero.

The pressure, net baryon number density, entropy density, energy density, and trace anomaly can be derived from the following thermodynamic relations:
\begin{align}
P(T,\mu_B) &= T \left( \frac{\partial \ln Z(T,\mu_B)}{\partial V} \right)_{T,\mu_B}, \\
n_B(T,\mu_B) &=\left( \frac{\partial P(T,\mu_B)}{\partial \mu_B} \right)_{T},   \\
s(T,\mu_B) &=  \left( \frac{\partial P(T,\mu_B)}{\partial T} \right)_{\mu_B},\\
\epsilon(T,\mu_B) &=Ts - P + \mu_B n_B,\\
\Delta(T,\mu_B) &=\epsilon - 3 P.
\label{eq:eos}
\end{align}
To predict the QCD equation of state at finite baryon chemical potential, we introduce the generalized susceptibilities $\chi_2^B$ and $\chi_4^B$ as one of the optimization objectives, defined by the following dimensionless derivatives:
\begin{align}
\chi_{i}^{B} = \frac{\partial{P(T,\hat{\mu}_B)/T^4}}{\partial{\hat{\mu}_B^i}}
\bigg |_{\hat{\mu}_B = 0}, \hat{\mu}_B = \mu_B/T.
\label{eq:chib}
\end{align}

The finally training objectives are composed of three components, one is the EoS constraint, which encompasses the entropy density, the trace anomaly. The second one is generalized susceptibilities $\chi_2^B$ and $\chi_4^B$. The last one is the masses constraint given by the hard thermal loop (HTL) result \cite{Haque:2024gva,Levai:1997yx}l:
\begin{align}
R_{g/q} = \frac{M_{g,T>2.5T_{cut}}}{M_{q,T>2.5T_{cut}}} = \sqrt{\frac{3}{2}\left(\frac{N_C}{3} + \frac{N_f}{6}\right)},
\label{eq:MCC}
\end{align} 
$T_{cut}$=0.150 GeV. Consequently, $m_{u/d}$ and $m_g$ can be expressed as:
\begin{align}
\mathcal{L}_1 = \left| R_{g/q} - \frac{3}{2} \right|.
\label{eq:LM1}
\end{align}
It is postulated that the $s$ quark at high temperatures satisfies:
\begin{align}
\mathcal{L}_2 = \left| \frac{m_s - m_{u/d}}{\overline{m}_s - \overline{m}_{u/d}} -1\right|.
\label{eq:LM2}
\end{align}
Here, the current quark masses are denoted by $\overline{m}_s \approx 95$~MeV and $\overline{m}_{u/d} \approx 5$~MeV. Additionally, it is found that the training results are better if the magnitudes of $s$ and $\Delta$ are similar. Therefore, we use $\Delta/T$ as a substitute for $\Delta$ in loss function. In the absence of numerical lattice QCD results at low temperatures, we use the Thermal-FIST package to generate the training data\cite{Vovchenko:2016rkn,Vovchenko:2019pjl}. The quasi-parton masses are then determined by minimizing the mean absolute error (MAE) between the network predictions and the lattice QCD data\cite{Bazavov:2017dus,Bollweg:2022rps}.

The total loss is thus given by,
\begin{align}
    \mathcal{L}(\theta_1, \theta_2, \theta_3)&=\left|s_{N N}-s_{\text {input }}\right|+\left|\frac{ \Delta_{N N}-\Delta_{\text {input }}}{T}\right| \\
    & +\left|\chi_{2, N N}^{B}-\chi_{2, \text { input }}^{B}\right|+\left|\chi_{4, N N}^{B}-\chi_{4, \text { input }}^{B}\right| \nonumber\\
    &+\mathcal{L}_{M C} \nonumber
\end{align}
where $\mathcal{L}_{M C} = \mathcal{L}_{1} + \mathcal{L}_{2}$ represents the mass constraint derived from high-temperature HTL calculations. Each trainable parameter $\theta$ is updated iteratively using SGD-like algorithms such as Adam, following the principle:
\begin{align}
    \theta \rightarrow \theta - \alpha{1\over n}\sum_i^n {\partial \mathcal{L}\over \partial \theta}
\end{align}
where $n$ is the number of training samples, $\alpha\sim 10^{-5}$ is the learning rate (a small positive number), and the ${\partial \mathcal{L}/ \partial \theta}$ is the gradient of the loss with respect to each trainable parameter. 

The ${\partial \mathcal{L}/ \partial \theta}$ is computed numerically using automatic differentiation (auto-diff), a highly efficient method for calculating derivatives with machine precision. The auto-diff is also employed to compute derivatives in $n_B$ (baryon number density), $s$ (entropy density) and $\chi_i^B$ (baryon susceptibility), automatically.

The momentum integration is implemented numerically using a 50-point Gaussian quadrature, facilitated by the PyTorch framework. This approach enables the network to propagate the loss to the trainable parameters of the mass models without requiring explicit computation of the derivatives of these integrals.

\begin{figure}[htp]
\centering
\includegraphics[width=0.48\textwidth]{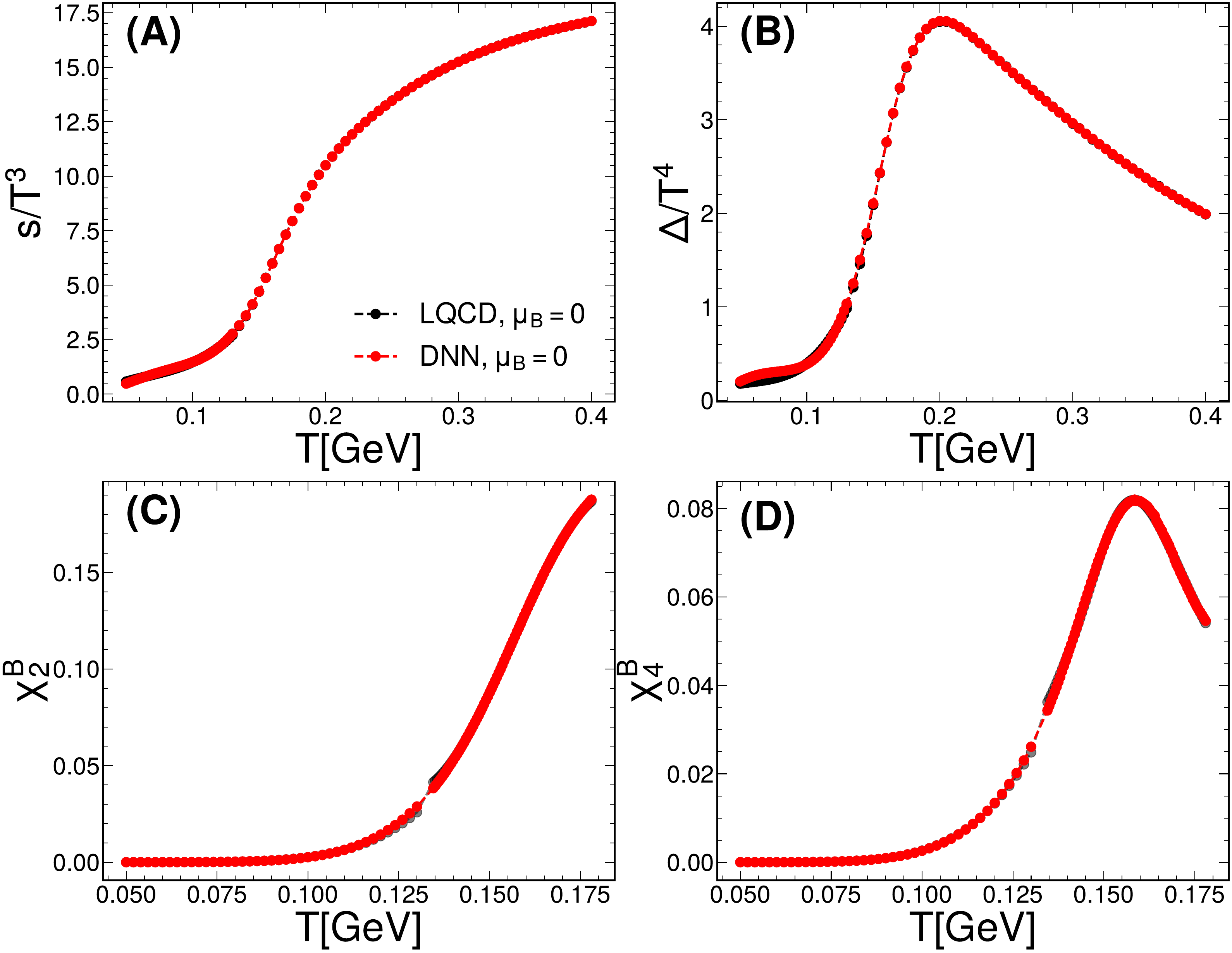}
\caption {\label{fig:train}(Color online)  The results obtained from training using Deep Learning methods are juxtaposed with those derived from lattice QCD (LQCD) and Hadron Resonance Gas (HRG) calculations at zero chemical potential, as detailed in \cite{HotQCD:2014kol,Vovchenko:2016rkn,Vovchenko:2019pjl}. Figures (A), (B), (C), and (D) depict the entropy density 
 $s/T^3$, trace anomaly $\Delta/T^3$, susceptibility $\chi_B^2$ and $\chi_B^4$ respectively, plotted against temperature. The red lines indicate the predictions from DNN, whereas the black lines signify the data from lattice QCD.}
\end{figure}

\begin{figure*}[htp]
\centering
\includegraphics[width=0.98\textwidth]{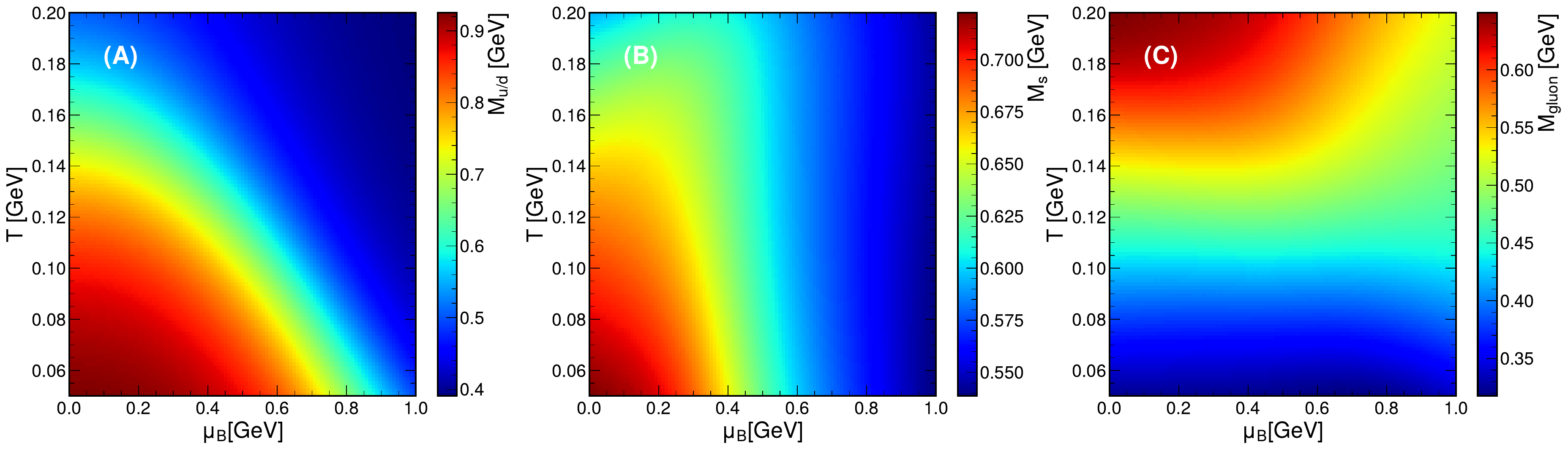}
\caption {\label{fig:mass}(Color online) The quasi-parton mass distribution at various $\mu_B$ and $T$ values is presented. From the left to the right panels are the distributions for quasi-$u/d$ quarks, quasi-$s$ quarks and gluons, respectively.}
\end{figure*}
\section{Results and discussion}
Fig.~\ref{fig:train} compares the values of $s/T^3$, $\Delta/T^4$, $\chi_2^B$, and $\chi_4^B$ with those obtained from lattice QCD (LQCD) and Hadron Resonance Gas (HRG) calculations. The results indicate that the quasi-parton masses reconstructed via DNN can precisely reproduce the QCD EoS across a wide temperature range using statistical mechanics formulations. Furthermore, DNN-learning quasi-parton masses can match the generalized susceptibilities $\chi_4^B$ and $\chi_2^B$ perfectly. The current EoS results cover the crossover transition region, which also indicates that our deep learning quasi-parton (DLQP) model is capable of capturing some of the phase transition properties of QCD.

\begin{figure*}[htp]
\centering
\includegraphics[width=0.98\textwidth]{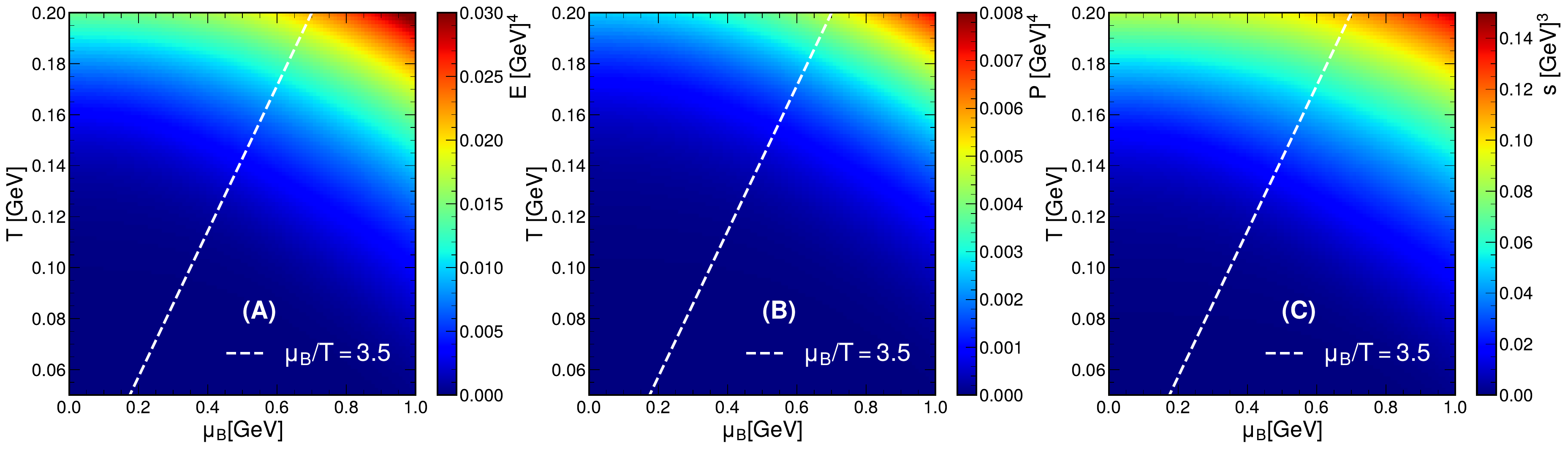}
\caption {\label{fig:finite_EOS}(Color online) The density distributions of energy density ($E$), pressure ($P$), and entropy density ($s/T^3$) as functions of $T$ and $\mu_B$are presented.}
\end{figure*}

\begin{figure}[htp]
\centering
\includegraphics[width=0.48\textwidth]{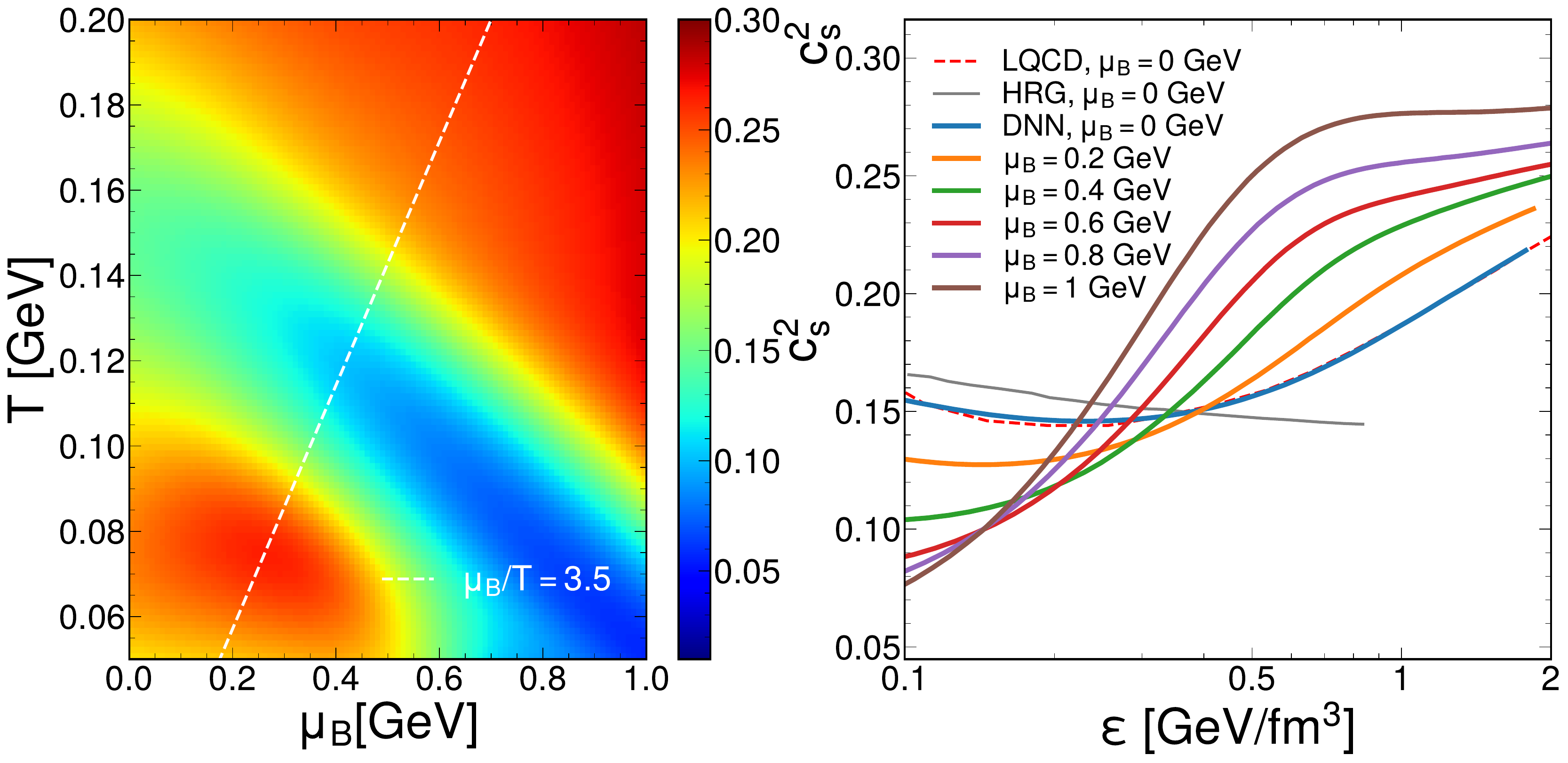}
\caption {\label{fig:cs_at_finite}(Color online) The left panel illustrates the distribution of the squared speed of sound $\c_s^2$ while the right panel depicts $c_s^2$ as a function of temperature at finite chemical potential, compared with lattice QCD calculations\cite{Plumari:2011mk}. The grey line represents the Hadron Resonance Gas (HRG) model calculation, and the red dashed line corresponds to results from the HotQCD collaboration\cite{HotQCD:2014kol}. The remaining colored lines denote different values of the baryon chemical potential.}
\end{figure}

Figure~\ref{fig:mass} shows the distribution of quasi-parton masses at different $T$ and $\mu_B$. It can be found that the $u/d$ quark and $s$ quark decrease as temperature increases at finite $\mu_B$, but the gluon increase as temperature increases. The results are different from other quasi-parton models based on the parameterized mass functions, which will introduce some priori bias\cite{Liu:2021dpm}. Once the quasi-parton mass is determined, we can easily compute various thermodynamic properties at different temperatures and baryon chemical potentials using Eq.~(\ref{eq:eos}). Figures~\ref{fig:finite_EOS} and \ref{fig:cs_at_finite} and  show the results for energy density $E$, pressure $P$, entropy density $s$ and speed of sound $c_s^2$ using quasi-parton masses. It can be found that EoS gradually increases as $T$ and $\mu_B$ increase. We also notice that $c_s^2$ first decreases to a minimum value and then increases at small $\mu_B$, which involves a crossover from hadron gas phase to QGP.

\begin{figure}[htp]
\centering
\includegraphics[width=0.48\textwidth]{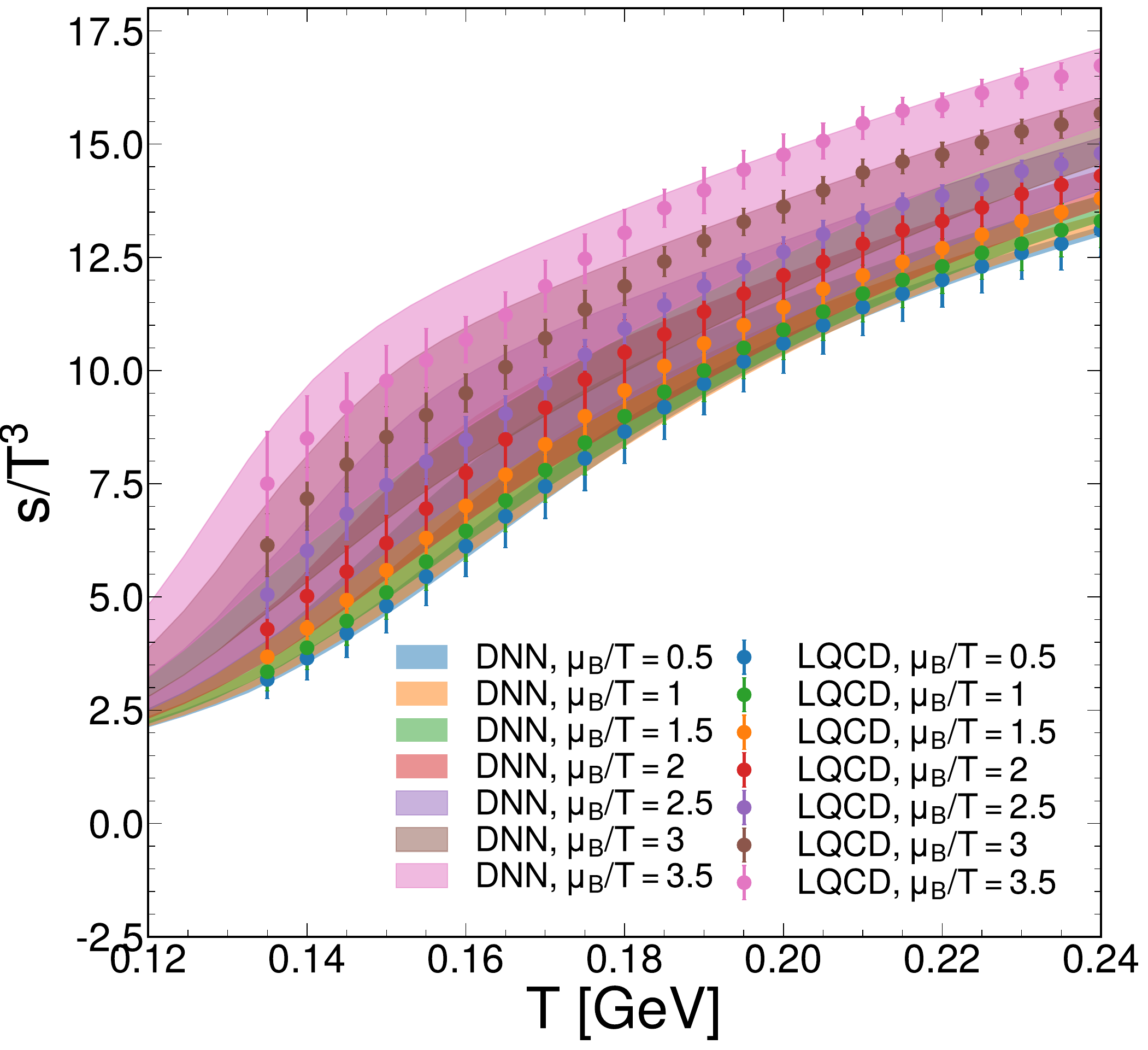}
\caption {\label{fig:s_at_finite}(Color online) The entropy density $s/T^3$ as a function of temperature at finite chemical potential, compared with the approximate expansion \cite{Borsanyi:2021sxv}. The color band illustrates the DNN-learned mass predictions, whereas the scatter plot represents the lattice QCD results \cite{Bollweg:2022rps,Bazavov:2017dus,Borsanyi:2021sxv}.}
\end{figure}

\begin{figure}[htp]
\centering
\includegraphics[width=0.48\textwidth]{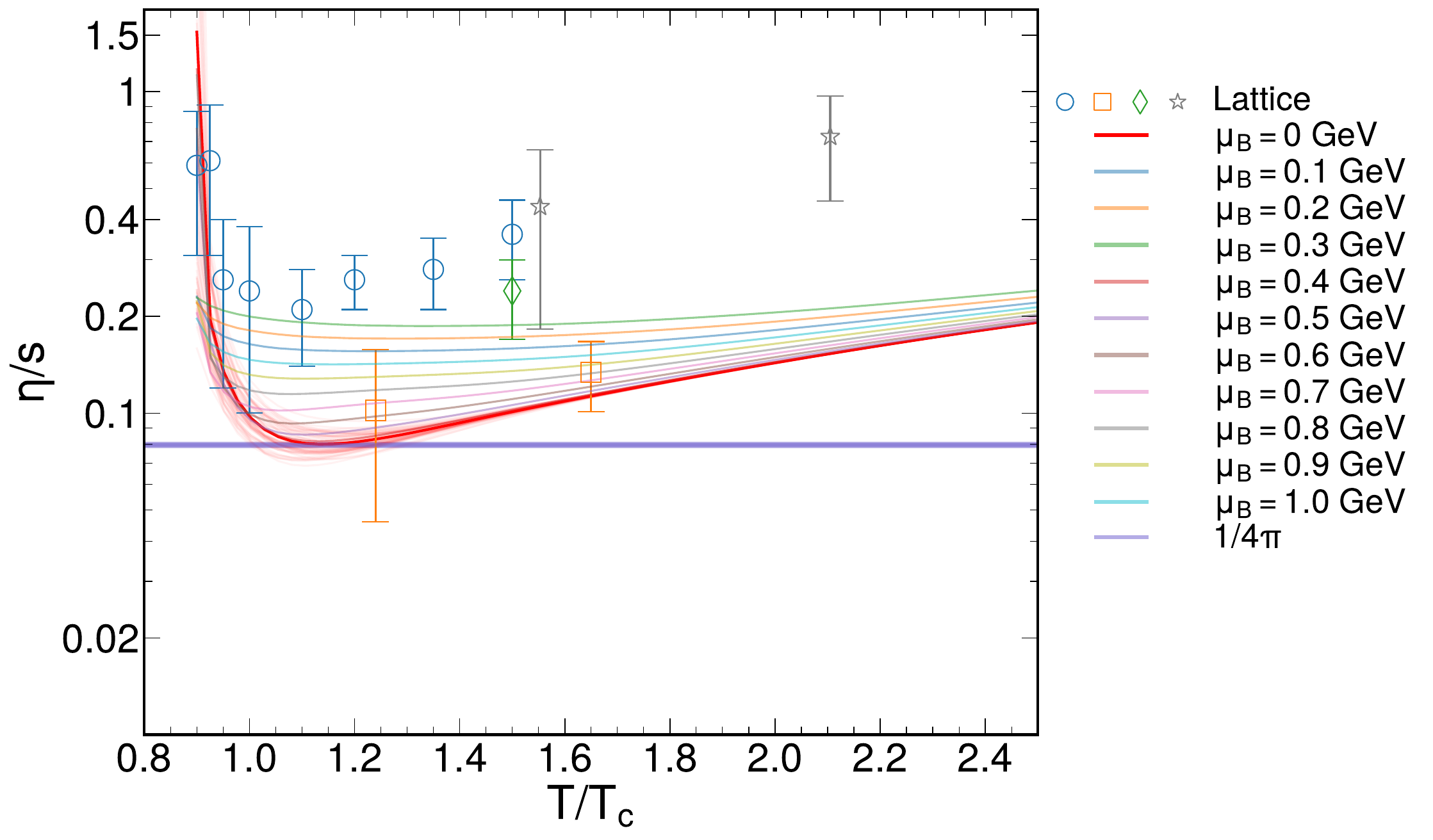}
\caption {\label{fig:eta_over_s_at_fin}(Color online) The temperature dependence of the shear viscosity to entropy density ratio $\eta/s$. The red curve represents our theoretical predictions obtained using the DLQP mass framework. The discrete data points correspond to lattice QCD calculations reported in \cite{Nakamura:2004sy,Astrakhantsev:2017nrs,Meyer:2007ic}. Additionally, we also present the 
$\mu_B$-dependent behavior of $\eta/s$ as a function of temperature.}
\end{figure}

\begin{figure}[htp]
\centering
\includegraphics[width=0.45\textwidth]{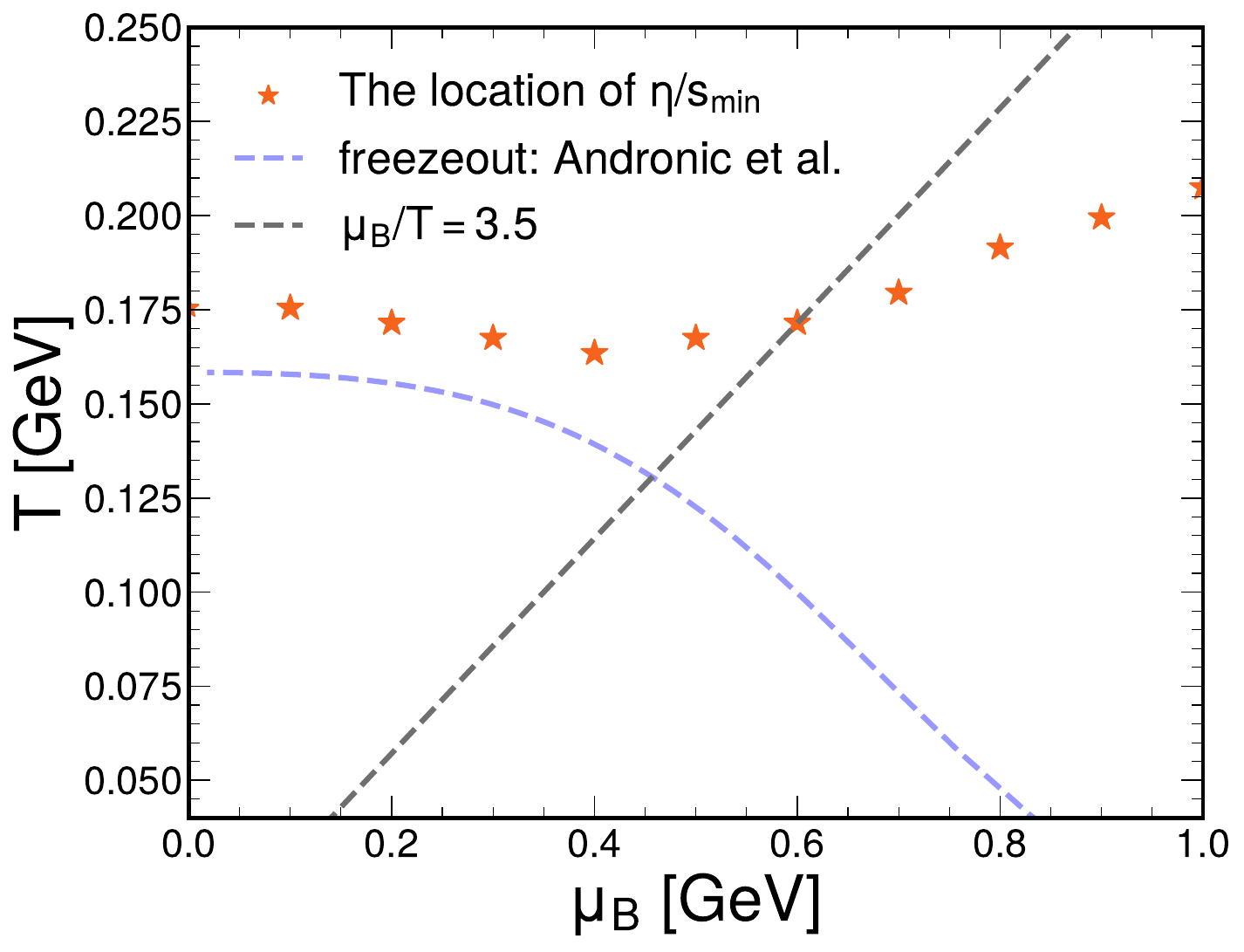}
\caption {\label{fig:min_eta}(Color online) The location of the minimum value of $\eta/s$ in the $T-\mu_B$ plane. The orange points represent the values computed using the masses learned by the DNN. The blue dashed line corresponds to the freeze-out result.\cite{Andronic:2009jd}.}
\end{figure}

Additionally, Figures~\ref{fig:cs_at_finite} and \ref{fig:s_at_finite} are the $s/T^3$ and $c_s^2$ as a function of temperature compared with LQCD calculations within $\mu_B/T=3.5$. It is evident that $s/T^3$ values increase as temperature rises, which is essentially cover lattice QCD calculations. The values of $c_s^2$ also increase as energy rises at different $\mu_B$. At zero chemical potential, we can perfectly reproduce the lattice QCD results. And as energy density $\epsilon$ increases, the trend of $c_s^2$ with respect to $\mu_B$ transitions from monotonic decrease to monotonic increase, which is likely due to a phase transition. These results not only validate the developed quasi-parton model for the study of the QCD EoS under finite temperature and baryon chemical potential conditions, but also furnish crucial EoS inputs for hydrodynamic simulations.

Using the relaxation time approximation approach \cite{Chen:2007xe,Plumari:2011mk,Plumari:2012ep}, we compute the QGP transport coefficient shear viscosity ($\eta$), which describes the properties of QGP fluid. Fig.~\ref{fig:eta_over_s_at_fin} presents the shear viscosity over entropy density ($\eta/s$) as a function of temperature at different $\mu_B$ values. The calculations using DLQP masses show that $\eta/s$ first rapidly drops to a minimum value around 175 MeV and then rises again as the temperature increases at zero $\mu_B$. This trend effectively aligns with results from lattice QCD calculations and agrees with the conclusions drawn in our previous work \cite{Li:2022ozl}, which suggests that the minimum value of $\eta/s$ should occur at a temperature greater than $T_c$. Furthermore, our findings indicate that the minimum value of $\eta/s$ increases with higher $\mu_B$, and the location of this minimum shifts accordingly.

Figure~\ref{fig:min_eta} shows the location of $\eta/s$ on the $T-\mu_B$ plane. The position of minimum value decreases as the $\mu_B$ increases within $\mu_B/T = 3.5$, but it will increase above $\mu_B/T=3.5$. And these positions are all located above the freeze-out temperature ($T_{ch}$), which indicates that the temperature required for the production of the hottest and densest QCD matter with the smallest shear viscosity must be greater than this $T_{ch}$. By comparing our current results with those obtained from previous methods, we demonstrate that our findings are reasonable at least within the range of $0\leq \mu_B/T\leq3.5$. Predictions beyond this range require additional physical constraints.

\section{Summary and outlook}
In this study, we have developed the deep-learning quasi-parton gas model to investigate the QCD EoS at finite baryon chemical potential. First, we introduce a novel methodology employing three neural networks to parameterize the quasi-gluon and quasi-quark masses. By utilizing data from the HRG model at low temperatures and lattice QCD results at high temperatures, we accurately reproduced the EoS using fundamental statistical mechanics principles. Crucially, our framework avoids any preconceived biases, relying exclusively on the powerful representational capabilities of neural networks and automatic differentiation techniques to address this variational problem.

Then, utilizing the temperature- and chemical potential-dependent masses, we calculate the entropy density, speed of sound, and shear viscosity as functions of temperature at various $\mu_B$. Our findings are largely consistent with those obtained from lattice QCD calculations through approximate expansion methods. Notably, at zero $\mu_B$, the minimum value of $\eta/s$ for the QGP occurs above the critical temperature ($T_c$). Additionally, as the baryon chemical potential increases, this minimum shifts to lower temperatures. Furthermore, the predictions of the EoS at finite $\mu_B$ provide essential inputs for relativistic hydrodynamic simulations.

Looking forward, we propose extending our deep-learning quasi-parton approach to include finite charge chemical potential ($\mu_Q$) and strangeness chemical potential ($\mu_S$), thereby facilitating a more comprehensive exploration of the EoS properties. Moreover, by incorporating the QCD critical point into our model, we aim to utilize this framework in the search for the QCD critical point. These future directions have the potential to significantly enhance our understanding of QCD phase transitions and their implications for high-energy physics.

\section{acknowledgments}
This work has been supported by the NSFC under grant Nos.\ 12075098, 12435009, 12225503 and 11935007. 
We gratefully acknowledge the extensive computing resources provided by the Nuclear Computing Center of Central China Normal University. 



\bibliography{ref}

\begin{thebibliography}{60}%
\makeatletter
\providecommand \@ifxundefined [1]{%
 \@ifx{#1\undefined}
}%
\providecommand \@ifnum [1]{%
 \ifnum #1\expandafter \@firstoftwo
 \else \expandafter \@secondoftwo
 \fi
}%
\providecommand \@ifx [1]{%
 \ifx #1\expandafter \@firstoftwo
 \else \expandafter \@secondoftwo
 \fi
}%
\providecommand \natexlab [1]{#1}%
\providecommand \enquote  [1]{``#1''}%
\providecommand \bibnamefont  [1]{#1}%
\providecommand \bibfnamefont [1]{#1}%
\providecommand \citenamefont [1]{#1}%
\providecommand \href@noop [0]{\@secondoftwo}%
\providecommand \href [0]{\begingroup \@sanitize@url \@href}%
\providecommand \@href[1]{\@@startlink{#1}\@@href}%
\providecommand \@@href[1]{\endgroup#1\@@endlink}%
\providecommand \@sanitize@url [0]{\catcode `\\12\catcode `\$12\catcode `\&12\catcode `\#12\catcode `\^12\catcode `\_12\catcode `\%12\relax}%
\providecommand \@@startlink[1]{}%
\providecommand \@@endlink[0]{}%
\providecommand \url  [0]{\begingroup\@sanitize@url \@url }%
\providecommand \@url [1]{\endgroup\@href {#1}{\urlprefix }}%
\providecommand \urlprefix  [0]{URL }%
\providecommand \Eprint [0]{\href }%
\providecommand \doibase [0]{http://dx.doi.org/}%
\providecommand \selectlanguage [0]{\@gobble}%
\providecommand \bibinfo  [0]{\@secondoftwo}%
\providecommand \bibfield  [0]{\@secondoftwo}%
\providecommand \translation [1]{[#1]}%
\providecommand \BibitemOpen [0]{}%
\providecommand \bibitemStop [0]{}%
\providecommand \bibitemNoStop [0]{.\EOS\space}%
\providecommand \EOS [0]{\spacefactor3000\relax}%
\providecommand \BibitemShut  [1]{\csname bibitem#1\endcsname}%
\let\auto@bib@innerbib\@empty
\bibitem [{\citenamefont {Akmal}\ \emph {et~al.}(1998)\citenamefont {Akmal}, \citenamefont {Pandharipande},\ and\ \citenamefont {Ravenhall}}]{Akmal:1998cf}%
  \BibitemOpen
  \bibfield  {author} {\bibinfo {author} {\bibfnamefont {A.}~\bibnamefont {Akmal}}, \bibinfo {author} {\bibfnamefont {V.~R.}\ \bibnamefont {Pandharipande}}, \ and\ \bibinfo {author} {\bibfnamefont {D.~G.}\ \bibnamefont {Ravenhall}},\ }\href {\doibase 10.1103/PhysRevC.58.1804} {\bibfield  {journal} {\bibinfo  {journal} {Phys. Rev. C}\ }\textbf {\bibinfo {volume} {58}},\ \bibinfo {pages} {1804} (\bibinfo {year} {1998})},\ \Eprint {http://arxiv.org/abs/nucl-th/9804027} {arXiv:nucl-th/9804027} \BibitemShut {NoStop}%
\bibitem [{\citenamefont {Lattimer}\ and\ \citenamefont {Prakash}(2001)}]{Lattimer:2000nx}%
  \BibitemOpen
  \bibfield  {author} {\bibinfo {author} {\bibfnamefont {J.~M.}\ \bibnamefont {Lattimer}}\ and\ \bibinfo {author} {\bibfnamefont {M.}~\bibnamefont {Prakash}},\ }\href {\doibase 10.1086/319702} {\bibfield  {journal} {\bibinfo  {journal} {Astrophys. J.}\ }\textbf {\bibinfo {volume} {550}},\ \bibinfo {pages} {426} (\bibinfo {year} {2001})},\ \Eprint {http://arxiv.org/abs/astro-ph/0002232} {arXiv:astro-ph/0002232} \BibitemShut {NoStop}%
\bibitem [{\citenamefont {Lattimer}(2012)}]{Lattimer:2012nd}%
  \BibitemOpen
  \bibfield  {author} {\bibinfo {author} {\bibfnamefont {J.~M.}\ \bibnamefont {Lattimer}},\ }\href {\doibase 10.1146/annurev-nucl-102711-095018} {\bibfield  {journal} {\bibinfo  {journal} {Ann. Rev. Nucl. Part. Sci.}\ }\textbf {\bibinfo {volume} {62}},\ \bibinfo {pages} {485} (\bibinfo {year} {2012})},\ \Eprint {http://arxiv.org/abs/1305.3510} {arXiv:1305.3510 [nucl-th]} \BibitemShut {NoStop}%
\bibitem [{\citenamefont {Braun-Munzinger}\ \emph {et~al.}(2016)\citenamefont {Braun-Munzinger}, \citenamefont {Koch}, \citenamefont {Sch\"afer},\ and\ \citenamefont {Stachel}}]{Braun-Munzinger:2015hba}%
  \BibitemOpen
  \bibfield  {author} {\bibinfo {author} {\bibfnamefont {P.}~\bibnamefont {Braun-Munzinger}}, \bibinfo {author} {\bibfnamefont {V.}~\bibnamefont {Koch}}, \bibinfo {author} {\bibfnamefont {T.}~\bibnamefont {Sch\"afer}}, \ and\ \bibinfo {author} {\bibfnamefont {J.}~\bibnamefont {Stachel}},\ }\href {\doibase 10.1016/j.physrep.2015.12.003} {\bibfield  {journal} {\bibinfo  {journal} {Phys. Rept.}\ }\textbf {\bibinfo {volume} {621}},\ \bibinfo {pages} {76} (\bibinfo {year} {2016})},\ \Eprint {http://arxiv.org/abs/1510.00442} {arXiv:1510.00442 [nucl-th]} \BibitemShut {NoStop}%
\bibitem [{\citenamefont {Adam}\ \emph {et~al.}(2017)\citenamefont {Adam} \emph {et~al.}}]{ALICE:2016fzo}%
  \BibitemOpen
  \bibfield  {author} {\bibinfo {author} {\bibfnamefont {J.}~\bibnamefont {Adam}} \emph {et~al.} (\bibinfo {collaboration} {ALICE}),\ }\href {\doibase 10.1038/nphys4111} {\bibfield  {journal} {\bibinfo  {journal} {Nature Phys.}\ }\textbf {\bibinfo {volume} {13}},\ \bibinfo {pages} {535} (\bibinfo {year} {2017})},\ \Eprint {http://arxiv.org/abs/1606.07424} {arXiv:1606.07424 [nucl-ex]} \BibitemShut {NoStop}%
\bibitem [{\citenamefont {Burgio}\ \emph {et~al.}(2021)\citenamefont {Burgio}, \citenamefont {Schulze}, \citenamefont {Vidana},\ and\ \citenamefont {Wei}}]{Burgio:2021vgk}%
  \BibitemOpen
  \bibfield  {author} {\bibinfo {author} {\bibfnamefont {G.~F.}\ \bibnamefont {Burgio}}, \bibinfo {author} {\bibfnamefont {H.~J.}\ \bibnamefont {Schulze}}, \bibinfo {author} {\bibfnamefont {I.}~\bibnamefont {Vidana}}, \ and\ \bibinfo {author} {\bibfnamefont {J.~B.}\ \bibnamefont {Wei}},\ }\href {\doibase 10.1016/j.ppnp.2021.103879} {\bibfield  {journal} {\bibinfo  {journal} {Prog. Part. Nucl. Phys.}\ }\textbf {\bibinfo {volume} {120}},\ \bibinfo {pages} {103879} (\bibinfo {year} {2021})},\ \Eprint {http://arxiv.org/abs/2105.03747} {arXiv:2105.03747 [nucl-th]} \BibitemShut {NoStop}%
\bibitem [{\citenamefont {Sorensen}\ \emph {et~al.}(2024)\citenamefont {Sorensen} \emph {et~al.}}]{Sorensen:2023zkk}%
  \BibitemOpen
  \bibfield  {author} {\bibinfo {author} {\bibfnamefont {A.}~\bibnamefont {Sorensen}} \emph {et~al.},\ }\href {\doibase 10.1016/j.ppnp.2023.104080} {\bibfield  {journal} {\bibinfo  {journal} {Prog. Part. Nucl. Phys.}\ }\textbf {\bibinfo {volume} {134}},\ \bibinfo {pages} {104080} (\bibinfo {year} {2024})},\ \Eprint {http://arxiv.org/abs/2301.13253} {arXiv:2301.13253 [nucl-th]} \BibitemShut {NoStop}%
\bibitem [{\citenamefont {Arslandok}\ \emph {et~al.}(2023)\citenamefont {Arslandok} \emph {et~al.}}]{Arslandok:2023utm}%
  \BibitemOpen
  \bibfield  {author} {\bibinfo {author} {\bibfnamefont {M.}~\bibnamefont {Arslandok}} \emph {et~al.},\ }\href@noop {} {\  (\bibinfo {year} {2023})},\ \Eprint {http://arxiv.org/abs/2303.17254} {arXiv:2303.17254 [nucl-ex]} \BibitemShut {NoStop}%
\bibitem [{\citenamefont {Kumar}\ \emph {et~al.}(2024)\citenamefont {Kumar} \emph {et~al.}}]{MUSES:2023hyz}%
  \BibitemOpen
  \bibfield  {author} {\bibinfo {author} {\bibfnamefont {R.}~\bibnamefont {Kumar}} \emph {et~al.} (\bibinfo {collaboration} {MUSES}),\ }\href {\doibase 10.1007/s41114-024-00049-6} {\bibfield  {journal} {\bibinfo  {journal} {Living Rev. Rel.}\ }\textbf {\bibinfo {volume} {27}},\ \bibinfo {pages} {3} (\bibinfo {year} {2024})},\ \Eprint {http://arxiv.org/abs/2303.17021} {arXiv:2303.17021 [nucl-th]} \BibitemShut {NoStop}%
\bibitem [{\citenamefont {Bazavov}\ \emph {et~al.}(2012{\natexlab{a}})\citenamefont {Bazavov} \emph {et~al.}}]{Bazavov:2011nk}%
  \BibitemOpen
  \bibfield  {author} {\bibinfo {author} {\bibfnamefont {A.}~\bibnamefont {Bazavov}} \emph {et~al.},\ }\href {\doibase 10.1103/PhysRevD.85.054503} {\bibfield  {journal} {\bibinfo  {journal} {Phys. Rev. D}\ }\textbf {\bibinfo {volume} {85}},\ \bibinfo {pages} {054503} (\bibinfo {year} {2012}{\natexlab{a}})},\ \Eprint {http://arxiv.org/abs/1111.1710} {arXiv:1111.1710 [hep-lat]} \BibitemShut {NoStop}%
\bibitem [{\citenamefont {Fukushima}\ and\ \citenamefont {Hatsuda}(2011)}]{Fukushima:2010bq}%
  \BibitemOpen
  \bibfield  {author} {\bibinfo {author} {\bibfnamefont {K.}~\bibnamefont {Fukushima}}\ and\ \bibinfo {author} {\bibfnamefont {T.}~\bibnamefont {Hatsuda}},\ }\href {\doibase 10.1088/0034-4885/74/1/014001} {\bibfield  {journal} {\bibinfo  {journal} {Rept. Prog. Phys.}\ }\textbf {\bibinfo {volume} {74}},\ \bibinfo {pages} {014001} (\bibinfo {year} {2011})},\ \Eprint {http://arxiv.org/abs/1005.4814} {arXiv:1005.4814 [hep-ph]} \BibitemShut {NoStop}%
\bibitem [{\citenamefont {Baym}\ \emph {et~al.}(2018)\citenamefont {Baym}, \citenamefont {Hatsuda}, \citenamefont {Kojo}, \citenamefont {Powell}, \citenamefont {Song},\ and\ \citenamefont {Takatsuka}}]{Baym:2017whm}%
  \BibitemOpen
  \bibfield  {author} {\bibinfo {author} {\bibfnamefont {G.}~\bibnamefont {Baym}}, \bibinfo {author} {\bibfnamefont {T.}~\bibnamefont {Hatsuda}}, \bibinfo {author} {\bibfnamefont {T.}~\bibnamefont {Kojo}}, \bibinfo {author} {\bibfnamefont {P.~D.}\ \bibnamefont {Powell}}, \bibinfo {author} {\bibfnamefont {Y.}~\bibnamefont {Song}}, \ and\ \bibinfo {author} {\bibfnamefont {T.}~\bibnamefont {Takatsuka}},\ }\href {\doibase 10.1088/1361-6633/aaae14} {\bibfield  {journal} {\bibinfo  {journal} {Rept. Prog. Phys.}\ }\textbf {\bibinfo {volume} {81}},\ \bibinfo {pages} {056902} (\bibinfo {year} {2018})},\ \Eprint {http://arxiv.org/abs/1707.04966} {arXiv:1707.04966 [astro-ph.HE]} \BibitemShut {NoStop}%
\bibitem [{\citenamefont {Andersen}\ \emph {et~al.}(2016)\citenamefont {Andersen}, \citenamefont {Naylor},\ and\ \citenamefont {Tranberg}}]{Andersen:2014xxa}%
  \BibitemOpen
  \bibfield  {author} {\bibinfo {author} {\bibfnamefont {J.~O.}\ \bibnamefont {Andersen}}, \bibinfo {author} {\bibfnamefont {W.~R.}\ \bibnamefont {Naylor}}, \ and\ \bibinfo {author} {\bibfnamefont {A.}~\bibnamefont {Tranberg}},\ }\href {\doibase 10.1103/RevModPhys.88.025001} {\bibfield  {journal} {\bibinfo  {journal} {Rev. Mod. Phys.}\ }\textbf {\bibinfo {volume} {88}},\ \bibinfo {pages} {025001} (\bibinfo {year} {2016})},\ \Eprint {http://arxiv.org/abs/1411.7176} {arXiv:1411.7176 [hep-ph]} \BibitemShut {NoStop}%
\bibitem [{\citenamefont {Fukushima}\ and\ \citenamefont {Sasaki}(2013)}]{Fukushima:2013rx}%
  \BibitemOpen
  \bibfield  {author} {\bibinfo {author} {\bibfnamefont {K.}~\bibnamefont {Fukushima}}\ and\ \bibinfo {author} {\bibfnamefont {C.}~\bibnamefont {Sasaki}},\ }\href {\doibase 10.1016/j.ppnp.2013.05.003} {\bibfield  {journal} {\bibinfo  {journal} {Prog. Part. Nucl. Phys.}\ }\textbf {\bibinfo {volume} {72}},\ \bibinfo {pages} {99} (\bibinfo {year} {2013})},\ \Eprint {http://arxiv.org/abs/1301.6377} {arXiv:1301.6377 [hep-ph]} \BibitemShut {NoStop}%
\bibitem [{\citenamefont {Bazavov}\ \emph {et~al.}(2014)\citenamefont {Bazavov} \emph {et~al.}}]{HotQCD:2014kol}%
  \BibitemOpen
  \bibfield  {author} {\bibinfo {author} {\bibfnamefont {A.}~\bibnamefont {Bazavov}} \emph {et~al.} (\bibinfo {collaboration} {HotQCD}),\ }\href {\doibase 10.1103/PhysRevD.90.094503} {\bibfield  {journal} {\bibinfo  {journal} {Phys. Rev. D}\ }\textbf {\bibinfo {volume} {90}},\ \bibinfo {pages} {094503} (\bibinfo {year} {2014})},\ \Eprint {http://arxiv.org/abs/1407.6387} {arXiv:1407.6387 [hep-lat]} \BibitemShut {NoStop}%
\bibitem [{\citenamefont {Bazavov}\ \emph {et~al.}(2019)\citenamefont {Bazavov} \emph {et~al.}}]{HotQCD:2018pds}%
  \BibitemOpen
  \bibfield  {author} {\bibinfo {author} {\bibfnamefont {A.}~\bibnamefont {Bazavov}} \emph {et~al.} (\bibinfo {collaboration} {HotQCD}),\ }\href {\doibase 10.1016/j.physletb.2019.05.013} {\bibfield  {journal} {\bibinfo  {journal} {Phys. Lett. B}\ }\textbf {\bibinfo {volume} {795}},\ \bibinfo {pages} {15} (\bibinfo {year} {2019})},\ \Eprint {http://arxiv.org/abs/1812.08235} {arXiv:1812.08235 [hep-lat]} \BibitemShut {NoStop}%
\bibitem [{\citenamefont {Bazavov}\ \emph {et~al.}(2017)\citenamefont {Bazavov} \emph {et~al.}}]{Bazavov:2017dus}%
  \BibitemOpen
  \bibfield  {author} {\bibinfo {author} {\bibfnamefont {A.}~\bibnamefont {Bazavov}} \emph {et~al.},\ }\href {\doibase 10.1103/PhysRevD.95.054504} {\bibfield  {journal} {\bibinfo  {journal} {Phys. Rev. D}\ }\textbf {\bibinfo {volume} {95}},\ \bibinfo {pages} {054504} (\bibinfo {year} {2017})},\ \Eprint {http://arxiv.org/abs/1701.04325} {arXiv:1701.04325 [hep-lat]} \BibitemShut {NoStop}%
\bibitem [{\citenamefont {Bazavov}\ \emph {et~al.}(2012{\natexlab{b}})\citenamefont {Bazavov} \emph {et~al.}}]{HotQCD:2012fhj}%
  \BibitemOpen
  \bibfield  {author} {\bibinfo {author} {\bibfnamefont {A.}~\bibnamefont {Bazavov}} \emph {et~al.} (\bibinfo {collaboration} {HotQCD}),\ }\href {\doibase 10.1103/PhysRevD.86.034509} {\bibfield  {journal} {\bibinfo  {journal} {Phys. Rev. D}\ }\textbf {\bibinfo {volume} {86}},\ \bibinfo {pages} {034509} (\bibinfo {year} {2012}{\natexlab{b}})},\ \Eprint {http://arxiv.org/abs/1203.0784} {arXiv:1203.0784 [hep-lat]} \BibitemShut {NoStop}%
\bibitem [{\citenamefont {Bazavov}\ \emph {et~al.}(2020)\citenamefont {Bazavov} \emph {et~al.}}]{Bazavov:2020bjn}%
  \BibitemOpen
  \bibfield  {author} {\bibinfo {author} {\bibfnamefont {A.}~\bibnamefont {Bazavov}} \emph {et~al.},\ }\href {\doibase 10.1103/PhysRevD.101.074502} {\bibfield  {journal} {\bibinfo  {journal} {Phys. Rev. D}\ }\textbf {\bibinfo {volume} {101}},\ \bibinfo {pages} {074502} (\bibinfo {year} {2020})},\ \Eprint {http://arxiv.org/abs/2001.08530} {arXiv:2001.08530 [hep-lat]} \BibitemShut {NoStop}%
\bibitem [{\citenamefont {de~Forcrand}\ and\ \citenamefont {Laliena}(2000)}]{deForcrand:1999fz}%
  \BibitemOpen
  \bibfield  {author} {\bibinfo {author} {\bibfnamefont {P.}~\bibnamefont {de~Forcrand}}\ and\ \bibinfo {author} {\bibfnamefont {V.}~\bibnamefont {Laliena}},\ }\href {\doibase 10.1016/S0920-5632(00)00316-9} {\bibfield  {journal} {\bibinfo  {journal} {Nucl. Phys. B Proc. Suppl.}\ }\textbf {\bibinfo {volume} {83}},\ \bibinfo {pages} {372} (\bibinfo {year} {2000})},\ \Eprint {http://arxiv.org/abs/hep-lat/9908015} {arXiv:hep-lat/9908015} \BibitemShut {NoStop}%
\bibitem [{\citenamefont {Engels}\ \emph {et~al.}(2000)\citenamefont {Engels}, \citenamefont {Kaczmarek}, \citenamefont {Karsch},\ and\ \citenamefont {Laermann}}]{Engels:1999id}%
  \BibitemOpen
  \bibfield  {author} {\bibinfo {author} {\bibfnamefont {J.}~\bibnamefont {Engels}}, \bibinfo {author} {\bibfnamefont {O.}~\bibnamefont {Kaczmarek}}, \bibinfo {author} {\bibfnamefont {F.}~\bibnamefont {Karsch}}, \ and\ \bibinfo {author} {\bibfnamefont {E.}~\bibnamefont {Laermann}},\ }\href {\doibase 10.1016/S0920-5632(00)91676-1} {\bibfield  {journal} {\bibinfo  {journal} {Nucl. Phys. B Proc. Suppl.}\ }\textbf {\bibinfo {volume} {83}},\ \bibinfo {pages} {369} (\bibinfo {year} {2000})},\ \Eprint {http://arxiv.org/abs/hep-lat/9908046} {arXiv:hep-lat/9908046} \BibitemShut {NoStop}%
\bibitem [{\citenamefont {Cox}\ \emph {et~al.}(2000)\citenamefont {Cox}, \citenamefont {Gattringer}, \citenamefont {Holland}, \citenamefont {Scarlet},\ and\ \citenamefont {Wiese}}]{Cox:1999nt}%
  \BibitemOpen
  \bibfield  {author} {\bibinfo {author} {\bibfnamefont {J.}~\bibnamefont {Cox}}, \bibinfo {author} {\bibfnamefont {C.}~\bibnamefont {Gattringer}}, \bibinfo {author} {\bibfnamefont {K.}~\bibnamefont {Holland}}, \bibinfo {author} {\bibfnamefont {B.}~\bibnamefont {Scarlet}}, \ and\ \bibinfo {author} {\bibfnamefont {U.~J.}\ \bibnamefont {Wiese}},\ }\href {\doibase 10.1016/S0920-5632(00)91804-8} {\bibfield  {journal} {\bibinfo  {journal} {Nucl. Phys. B Proc. Suppl.}\ }\textbf {\bibinfo {volume} {83}},\ \bibinfo {pages} {777} (\bibinfo {year} {2000})},\ \Eprint {http://arxiv.org/abs/hep-lat/9909119} {arXiv:hep-lat/9909119} \BibitemShut {NoStop}%
\bibitem [{\citenamefont {Bollweg}\ \emph {et~al.}(2022)\citenamefont {Bollweg}, \citenamefont {Goswami}, \citenamefont {Kaczmarek}, \citenamefont {Karsch}, \citenamefont {Mukherjee}, \citenamefont {Petreczky}, \citenamefont {Schmidt},\ and\ \citenamefont {Scior}}]{Bollweg:2022rps}%
  \BibitemOpen
  \bibfield  {author} {\bibinfo {author} {\bibfnamefont {D.}~\bibnamefont {Bollweg}}, \bibinfo {author} {\bibfnamefont {J.}~\bibnamefont {Goswami}}, \bibinfo {author} {\bibfnamefont {O.}~\bibnamefont {Kaczmarek}}, \bibinfo {author} {\bibfnamefont {F.}~\bibnamefont {Karsch}}, \bibinfo {author} {\bibfnamefont {S.}~\bibnamefont {Mukherjee}}, \bibinfo {author} {\bibfnamefont {P.}~\bibnamefont {Petreczky}}, \bibinfo {author} {\bibfnamefont {C.}~\bibnamefont {Schmidt}}, \ and\ \bibinfo {author} {\bibfnamefont {P.}~\bibnamefont {Scior}} (\bibinfo {collaboration} {HotQCD}),\ }\href {\doibase 10.1103/PhysRevD.105.074511} {\bibfield  {journal} {\bibinfo  {journal} {Phys. Rev. D}\ }\textbf {\bibinfo {volume} {105}},\ \bibinfo {pages} {074511} (\bibinfo {year} {2022})},\ \Eprint {http://arxiv.org/abs/2202.09184} {arXiv:2202.09184 [hep-lat]} \BibitemShut {NoStop}%
\bibitem [{\citenamefont {Bors\'anyi}\ \emph {et~al.}(2021)\citenamefont {Bors\'anyi}, \citenamefont {Fodor}, \citenamefont {Guenther}, \citenamefont {Kara}, \citenamefont {Katz}, \citenamefont {Parotto}, \citenamefont {P\'asztor}, \citenamefont {Ratti},\ and\ \citenamefont {Szab\'o}}]{Borsanyi:2021sxv}%
  \BibitemOpen
  \bibfield  {author} {\bibinfo {author} {\bibfnamefont {S.}~\bibnamefont {Bors\'anyi}}, \bibinfo {author} {\bibfnamefont {Z.}~\bibnamefont {Fodor}}, \bibinfo {author} {\bibfnamefont {J.~N.}\ \bibnamefont {Guenther}}, \bibinfo {author} {\bibfnamefont {R.}~\bibnamefont {Kara}}, \bibinfo {author} {\bibfnamefont {S.~D.}\ \bibnamefont {Katz}}, \bibinfo {author} {\bibfnamefont {P.}~\bibnamefont {Parotto}}, \bibinfo {author} {\bibfnamefont {A.}~\bibnamefont {P\'asztor}}, \bibinfo {author} {\bibfnamefont {C.}~\bibnamefont {Ratti}}, \ and\ \bibinfo {author} {\bibfnamefont {K.~K.}\ \bibnamefont {Szab\'o}},\ }\href {\doibase 10.1103/PhysRevLett.126.232001} {\bibfield  {journal} {\bibinfo  {journal} {Phys. Rev. Lett.}\ }\textbf {\bibinfo {volume} {126}},\ \bibinfo {pages} {232001} (\bibinfo {year} {2021})},\ \Eprint {http://arxiv.org/abs/2102.06660} {arXiv:2102.06660 [hep-lat]} \BibitemShut {NoStop}%
\bibitem [{\citenamefont {Monnai}\ \emph {et~al.}(2019)\citenamefont {Monnai}, \citenamefont {Schenke},\ and\ \citenamefont {Shen}}]{Monnai:2019hkn}%
  \BibitemOpen
  \bibfield  {author} {\bibinfo {author} {\bibfnamefont {A.}~\bibnamefont {Monnai}}, \bibinfo {author} {\bibfnamefont {B.}~\bibnamefont {Schenke}}, \ and\ \bibinfo {author} {\bibfnamefont {C.}~\bibnamefont {Shen}},\ }\href {\doibase 10.1103/PhysRevC.100.024907} {\bibfield  {journal} {\bibinfo  {journal} {Phys. Rev. C}\ }\textbf {\bibinfo {volume} {100}},\ \bibinfo {pages} {024907} (\bibinfo {year} {2019})},\ \Eprint {http://arxiv.org/abs/1902.05095} {arXiv:1902.05095 [nucl-th]} \BibitemShut {NoStop}%
\bibitem [{\citenamefont {Vovchenko}\ \emph {et~al.}(2017)\citenamefont {Vovchenko}, \citenamefont {Gorenstein},\ and\ \citenamefont {Stoecker}}]{Vovchenko:2016rkn}%
  \BibitemOpen
  \bibfield  {author} {\bibinfo {author} {\bibfnamefont {V.}~\bibnamefont {Vovchenko}}, \bibinfo {author} {\bibfnamefont {M.~I.}\ \bibnamefont {Gorenstein}}, \ and\ \bibinfo {author} {\bibfnamefont {H.}~\bibnamefont {Stoecker}},\ }\href {\doibase 10.1103/PhysRevLett.118.182301} {\bibfield  {journal} {\bibinfo  {journal} {Phys. Rev. Lett.}\ }\textbf {\bibinfo {volume} {118}},\ \bibinfo {pages} {182301} (\bibinfo {year} {2017})},\ \Eprint {http://arxiv.org/abs/1609.03975} {arXiv:1609.03975 [hep-ph]} \BibitemShut {NoStop}%
\bibitem [{\citenamefont {Vovchenko}\ and\ \citenamefont {Stoecker}(2019)}]{Vovchenko:2019pjl}%
  \BibitemOpen
  \bibfield  {author} {\bibinfo {author} {\bibfnamefont {V.}~\bibnamefont {Vovchenko}}\ and\ \bibinfo {author} {\bibfnamefont {H.}~\bibnamefont {Stoecker}},\ }\href {\doibase 10.1016/j.cpc.2019.06.024} {\bibfield  {journal} {\bibinfo  {journal} {Comput. Phys. Commun.}\ }\textbf {\bibinfo {volume} {244}},\ \bibinfo {pages} {295} (\bibinfo {year} {2019})},\ \Eprint {http://arxiv.org/abs/1901.05249} {arXiv:1901.05249 [nucl-th]} \BibitemShut {NoStop}%
\bibitem [{\citenamefont {Monnai}\ \emph {et~al.}(2024)\citenamefont {Monnai}, \citenamefont {Pihan}, \citenamefont {Schenke},\ and\ \citenamefont {Shen}}]{Monnai:2024pvy}%
  \BibitemOpen
  \bibfield  {author} {\bibinfo {author} {\bibfnamefont {A.}~\bibnamefont {Monnai}}, \bibinfo {author} {\bibfnamefont {G.}~\bibnamefont {Pihan}}, \bibinfo {author} {\bibfnamefont {B.}~\bibnamefont {Schenke}}, \ and\ \bibinfo {author} {\bibfnamefont {C.}~\bibnamefont {Shen}},\ }\href {\doibase 10.1103/PhysRevC.110.044905} {\bibfield  {journal} {\bibinfo  {journal} {Phys. Rev. C}\ }\textbf {\bibinfo {volume} {110}},\ \bibinfo {pages} {044905} (\bibinfo {year} {2024})},\ \Eprint {http://arxiv.org/abs/2406.11610} {arXiv:2406.11610 [nucl-th]} \BibitemShut {NoStop}%
\bibitem [{\citenamefont {Raissi}\ \emph {et~al.}(2019)\citenamefont {Raissi}, \citenamefont {Perdikaris},\ and\ \citenamefont {Karniadakis}}]{2018PINN}%
  \BibitemOpen
  \bibfield  {author} {\bibinfo {author} {\bibfnamefont {M.}~\bibnamefont {Raissi}}, \bibinfo {author} {\bibfnamefont {P.}~\bibnamefont {Perdikaris}}, \ and\ \bibinfo {author} {\bibfnamefont {G.~E.}\ \bibnamefont {Karniadakis}},\ }\href {\doibase https://doi.org/10.1016/j.jcp.2018.10.045} {\bibfield  {journal} {\bibinfo  {journal} {Journal of Computational Physics}\ }\textbf {\bibinfo {volume} {378}},\ \bibinfo {pages} {686} (\bibinfo {year} {2019})}\BibitemShut {NoStop}%
\bibitem [{\citenamefont {Lagaris}\ \emph {et~al.}(1998)\citenamefont {Lagaris}, \citenamefont {Likas},\ and\ \citenamefont {Fotiadis}}]{pde712178}%
  \BibitemOpen
  \bibfield  {author} {\bibinfo {author} {\bibfnamefont {I.}~\bibnamefont {Lagaris}}, \bibinfo {author} {\bibfnamefont {A.}~\bibnamefont {Likas}}, \ and\ \bibinfo {author} {\bibfnamefont {D.}~\bibnamefont {Fotiadis}},\ }\href {\doibase 10.1109/72.712178} {\bibfield  {journal} {\bibinfo  {journal} {IEEE Transactions on Neural Networks}\ }\textbf {\bibinfo {volume} {9}},\ \bibinfo {pages} {987} (\bibinfo {year} {1998})}\BibitemShut {NoStop}%
\bibitem [{\citenamefont {KHOO}\ \emph {et~al.}(2021)\citenamefont {KHOO}, \citenamefont {LU},\ and\ \citenamefont {YING}}]{khoo_lu_ying_2021}%
  \BibitemOpen
  \bibfield  {author} {\bibinfo {author} {\bibfnamefont {Y.}~\bibnamefont {KHOO}}, \bibinfo {author} {\bibfnamefont {J.}~\bibnamefont {LU}}, \ and\ \bibinfo {author} {\bibfnamefont {L.}~\bibnamefont {YING}},\ }\href {\doibase 10.1017/S0956792520000182} {\bibfield  {journal} {\bibinfo  {journal} {European Journal of Applied Mathematics}\ }\textbf {\bibinfo {volume} {32}},\ \bibinfo {pages} {421–435} (\bibinfo {year} {2021})}\BibitemShut {NoStop}%
\bibitem [{\citenamefont {Ramuhalli}\ \emph {et~al.}(2005)\citenamefont {Ramuhalli}, \citenamefont {Udpa},\ and\ \citenamefont {Udpa}}]{pde1528518}%
  \BibitemOpen
  \bibfield  {author} {\bibinfo {author} {\bibfnamefont {P.}~\bibnamefont {Ramuhalli}}, \bibinfo {author} {\bibfnamefont {L.}~\bibnamefont {Udpa}}, \ and\ \bibinfo {author} {\bibfnamefont {S.}~\bibnamefont {Udpa}},\ }\href {\doibase 10.1109/TNN.2005.857945} {\bibfield  {journal} {\bibinfo  {journal} {IEEE Transactions on Neural Networks}\ }\textbf {\bibinfo {volume} {16}},\ \bibinfo {pages} {1381} (\bibinfo {year} {2005})}\BibitemShut {NoStop}%
\bibitem [{\citenamefont {Sirignano}\ and\ \citenamefont {Spiliopoulos}(2017)}]{2017DGM}%
  \BibitemOpen
  \bibfield  {author} {\bibinfo {author} {\bibfnamefont {J.}~\bibnamefont {Sirignano}}\ and\ \bibinfo {author} {\bibfnamefont {K.}~\bibnamefont {Spiliopoulos}},\ }\href {\doibase https://doi.org/10.1016/j.jcp.2018.08.029} {\bibfield  {journal} {\bibinfo  {journal} {Journal of Computational Physics}\ }\textbf {\bibinfo {volume} {375}} (\bibinfo {year} {2017}),\ https://doi.org/10.1016/j.jcp.2018.08.029}\BibitemShut {NoStop}%
\bibitem [{\citenamefont {Weinan}\ and\ \citenamefont {Yu}(2018)}]{2018The}%
  \BibitemOpen
  \bibfield  {author} {\bibinfo {author} {\bibfnamefont {E.}~\bibnamefont {Weinan}}\ and\ \bibinfo {author} {\bibfnamefont {B.}~\bibnamefont {Yu}},\ }\href {\doibase https://doi.org/10.1007/s40304-018-0127-z} {\bibfield  {journal} {\bibinfo  {journal} {Communications in Mathematics \& Statistics}\ }\textbf {\bibinfo {volume} {6}},\ \bibinfo {pages} {1} (\bibinfo {year} {2018})}\BibitemShut {NoStop}%
\bibitem [{\citenamefont {Liao}\ and\ \citenamefont {Ming}(2021)}]{2021Deep}%
  \BibitemOpen
  \bibfield  {author} {\bibinfo {author} {\bibfnamefont {Y.}~\bibnamefont {Liao}}\ and\ \bibinfo {author} {\bibfnamefont {P.}~\bibnamefont {Ming}},\ }\href {\doibase https://doi.org/10.4208/cicp.OA-2020-0219} {\bibfield  {journal} {\bibinfo  {journal} {Communications in Computational Physics}\ }\textbf {\bibinfo {volume} {29}},\ \bibinfo {pages} {1365} (\bibinfo {year} {2021})}\BibitemShut {NoStop}%
\bibitem [{\citenamefont {M.~Raissi}\ and\ \citenamefont {Karniadakis}(2019)}]{RAISSI2019686}%
  \BibitemOpen
  \bibfield  {author} {\bibinfo {author} {\bibfnamefont {P.~P.}\ \bibnamefont {M.~Raissi}}\ and\ \bibinfo {author} {\bibfnamefont {G.}~\bibnamefont {Karniadakis}},\ }\href {\doibase https://doi.org/10.1016/j.jcp.2018.10.045} {\bibfield  {journal} {\bibinfo  {journal} {Journal of Computational Physics}\ }\textbf {\bibinfo {volume} {378}},\ \bibinfo {pages} {686} (\bibinfo {year} {2019})}\BibitemShut {NoStop}%
\bibitem [{\citenamefont {Soma}\ \emph {et~al.}(2022)\citenamefont {Soma}, \citenamefont {Wang}, \citenamefont {Shi}, \citenamefont {St\"ocker},\ and\ \citenamefont {Zhou}}]{Soma:2022vbb}%
  \BibitemOpen
  \bibfield  {author} {\bibinfo {author} {\bibfnamefont {S.}~\bibnamefont {Soma}}, \bibinfo {author} {\bibfnamefont {L.}~\bibnamefont {Wang}}, \bibinfo {author} {\bibfnamefont {S.}~\bibnamefont {Shi}}, \bibinfo {author} {\bibfnamefont {H.}~\bibnamefont {St\"ocker}}, \ and\ \bibinfo {author} {\bibfnamefont {K.}~\bibnamefont {Zhou}},\ }\href@noop {} {\  (\bibinfo {year} {2022})},\ \Eprint {http://arxiv.org/abs/2209.08883} {arXiv:2209.08883 [astro-ph.HE]} \BibitemShut {NoStop}%
\bibitem [{\citenamefont {Zhou}\ \emph {et~al.}(2024)\citenamefont {Zhou}, \citenamefont {Wang}, \citenamefont {Pang},\ and\ \citenamefont {Shi}}]{Zhou:2023pti}%
  \BibitemOpen
  \bibfield  {author} {\bibinfo {author} {\bibfnamefont {K.}~\bibnamefont {Zhou}}, \bibinfo {author} {\bibfnamefont {L.}~\bibnamefont {Wang}}, \bibinfo {author} {\bibfnamefont {L.-G.}\ \bibnamefont {Pang}}, \ and\ \bibinfo {author} {\bibfnamefont {S.}~\bibnamefont {Shi}},\ }\href {\doibase 10.1016/j.ppnp.2023.104084} {\bibfield  {journal} {\bibinfo  {journal} {Prog. Part. Nucl. Phys.}\ }\textbf {\bibinfo {volume} {135}},\ \bibinfo {pages} {104084} (\bibinfo {year} {2024})},\ \Eprint {http://arxiv.org/abs/2303.15136} {arXiv:2303.15136 [hep-ph]} \BibitemShut {NoStop}%
\bibitem [{\citenamefont {Pang}\ \emph {et~al.}(2018)\citenamefont {Pang}, \citenamefont {Zhou}, \citenamefont {Su}, \citenamefont {Petersen}, \citenamefont {St\"ocker},\ and\ \citenamefont {Wang}}]{Pang:2016vdc}%
  \BibitemOpen
  \bibfield  {author} {\bibinfo {author} {\bibfnamefont {L.-G.}\ \bibnamefont {Pang}}, \bibinfo {author} {\bibfnamefont {K.}~\bibnamefont {Zhou}}, \bibinfo {author} {\bibfnamefont {N.}~\bibnamefont {Su}}, \bibinfo {author} {\bibfnamefont {H.}~\bibnamefont {Petersen}}, \bibinfo {author} {\bibfnamefont {H.}~\bibnamefont {St\"ocker}}, \ and\ \bibinfo {author} {\bibfnamefont {X.-N.}\ \bibnamefont {Wang}},\ }\href {\doibase 10.1038/s41467-017-02726-3} {\bibfield  {journal} {\bibinfo  {journal} {Nature Commun.}\ }\textbf {\bibinfo {volume} {9}},\ \bibinfo {pages} {210} (\bibinfo {year} {2018})},\ \Eprint {http://arxiv.org/abs/1612.04262} {arXiv:1612.04262 [hep-ph]} \BibitemShut {NoStop}%
\bibitem [{\citenamefont {Boehnlein}\ \emph {et~al.}(2022)\citenamefont {Boehnlein} \emph {et~al.}}]{Boehnlein:2021eym}%
  \BibitemOpen
  \bibfield  {author} {\bibinfo {author} {\bibfnamefont {A.}~\bibnamefont {Boehnlein}} \emph {et~al.},\ }\href {\doibase 10.1103/RevModPhys.94.031003} {\bibfield  {journal} {\bibinfo  {journal} {Rev. Mod. Phys.}\ }\textbf {\bibinfo {volume} {94}},\ \bibinfo {pages} {031003} (\bibinfo {year} {2022})},\ \Eprint {http://arxiv.org/abs/2112.02309} {arXiv:2112.02309 [nucl-th]} \BibitemShut {NoStop}%
\bibitem [{\citenamefont {Aarts}\ \emph {et~al.}(2025)\citenamefont {Aarts}, \citenamefont {Fukushima}, \citenamefont {Hatsuda}, \citenamefont {Ipp}, \citenamefont {Shi}, \citenamefont {Wang},\ and\ \citenamefont {Zhou}}]{Aarts2025}%
  \BibitemOpen
  \bibfield  {author} {\bibinfo {author} {\bibfnamefont {G.}~\bibnamefont {Aarts}}, \bibinfo {author} {\bibfnamefont {K.}~\bibnamefont {Fukushima}}, \bibinfo {author} {\bibfnamefont {T.}~\bibnamefont {Hatsuda}}, \bibinfo {author} {\bibfnamefont {A.}~\bibnamefont {Ipp}}, \bibinfo {author} {\bibfnamefont {S.}~\bibnamefont {Shi}}, \bibinfo {author} {\bibfnamefont {L.}~\bibnamefont {Wang}}, \ and\ \bibinfo {author} {\bibfnamefont {K.}~\bibnamefont {Zhou}},\ }\href {\doibase 10.1038/s42254-024-00798-x} {\bibfield  {journal} {\bibinfo  {journal} {Nature Reviews Physics}\ }\textbf {\bibinfo {volume} {7}},\ \bibinfo {pages} {1} (\bibinfo {year} {2025})}\BibitemShut {NoStop}%
\bibitem [{\citenamefont {Pang}(2024)}]{Pang:2024kid}%
  \BibitemOpen
  \bibfield  {author} {\bibinfo {author} {\bibfnamefont {L.-G.}\ \bibnamefont {Pang}},\ }\href {\doibase 10.1142/S0218301324300091} {\bibfield  {journal} {\bibinfo  {journal} {Int. J. Mod. Phys. E}\ }\textbf {\bibinfo {volume} {33}},\ \bibinfo {pages} {2430009} (\bibinfo {year} {2024})}\BibitemShut {NoStop}%
\bibitem [{\citenamefont {Hornik}\ \emph {et~al.}(1989)\citenamefont {Hornik}, \citenamefont {Stinchcombe},\ and\ \citenamefont {White}}]{Hornik1989MultilayerFN}%
  \BibitemOpen
  \bibfield  {author} {\bibinfo {author} {\bibfnamefont {K.}~\bibnamefont {Hornik}}, \bibinfo {author} {\bibfnamefont {M.~B.}\ \bibnamefont {Stinchcombe}}, \ and\ \bibinfo {author} {\bibfnamefont {H.~L.}\ \bibnamefont {White}},\ }\href@noop {} {\bibfield  {journal} {\bibinfo  {journal} {Neural Networks}\ }\textbf {\bibinfo {volume} {2}},\ \bibinfo {pages} {359} (\bibinfo {year} {1989})}\BibitemShut {NoStop}%
\bibitem [{\citenamefont {Baydin}\ \emph {et~al.}(2018)\citenamefont {Baydin}, \citenamefont {Pearlmutter}, \citenamefont {Radul},\ and\ \citenamefont {Siskind}}]{JMLR:v18:17-468}%
  \BibitemOpen
  \bibfield  {author} {\bibinfo {author} {\bibfnamefont {A.~G.}\ \bibnamefont {Baydin}}, \bibinfo {author} {\bibfnamefont {B.~A.}\ \bibnamefont {Pearlmutter}}, \bibinfo {author} {\bibfnamefont {A.~A.}\ \bibnamefont {Radul}}, \ and\ \bibinfo {author} {\bibfnamefont {J.~M.}\ \bibnamefont {Siskind}},\ }\href {http://jmlr.org/papers/v18/17-468.html} {\bibfield  {journal} {\bibinfo  {journal} {Journal of Machine Learning Research}\ }\textbf {\bibinfo {volume} {18}},\ \bibinfo {pages} {1} (\bibinfo {year} {2018})}\BibitemShut {NoStop}%
\bibitem [{\citenamefont {Mattuck}(1976)}]{1976A}%
  \BibitemOpen
  \bibfield  {author} {\bibinfo {author} {\bibfnamefont {R.}~\bibnamefont {Mattuck}},\ }\href@noop {} {\bibfield  {journal} {\bibinfo  {journal} {Osborne McGraw-Hill}\ } (\bibinfo {year} {1976})}\BibitemShut {NoStop}%
\bibitem [{\citenamefont {Liu}\ \emph {et~al.}(2022)\citenamefont {Liu}, \citenamefont {Xing}, \citenamefont {Wu}, \citenamefont {Qin}, \citenamefont {Cao},\ and\ \citenamefont {Wang}}]{Liu:2021dpm}%
  \BibitemOpen
  \bibfield  {author} {\bibinfo {author} {\bibfnamefont {F.-L.}\ \bibnamefont {Liu}}, \bibinfo {author} {\bibfnamefont {W.-J.}\ \bibnamefont {Xing}}, \bibinfo {author} {\bibfnamefont {X.-Y.}\ \bibnamefont {Wu}}, \bibinfo {author} {\bibfnamefont {G.-Y.}\ \bibnamefont {Qin}}, \bibinfo {author} {\bibfnamefont {S.}~\bibnamefont {Cao}}, \ and\ \bibinfo {author} {\bibfnamefont {X.-N.}\ \bibnamefont {Wang}},\ }\href {\doibase 10.1140/epjc/s10052-022-10308-x} {\bibfield  {journal} {\bibinfo  {journal} {Eur. Phys. J. C}\ }\textbf {\bibinfo {volume} {82}},\ \bibinfo {pages} {350} (\bibinfo {year} {2022})},\ \Eprint {http://arxiv.org/abs/2107.11713} {arXiv:2107.11713 [hep-ph]} \BibitemShut {NoStop}%
\bibitem [{\citenamefont {Soloveva}\ \emph {et~al.}(2022)\citenamefont {Soloveva}, \citenamefont {Aichelin},\ and\ \citenamefont {Bratkovskaya}}]{Soloveva:2021quj}%
  \BibitemOpen
  \bibfield  {author} {\bibinfo {author} {\bibfnamefont {O.}~\bibnamefont {Soloveva}}, \bibinfo {author} {\bibfnamefont {J.}~\bibnamefont {Aichelin}}, \ and\ \bibinfo {author} {\bibfnamefont {E.}~\bibnamefont {Bratkovskaya}},\ }\href {\doibase 10.1103/PhysRevD.105.054011} {\bibfield  {journal} {\bibinfo  {journal} {Phys. Rev. D}\ }\textbf {\bibinfo {volume} {105}},\ \bibinfo {pages} {054011} (\bibinfo {year} {2022})},\ \Eprint {http://arxiv.org/abs/2108.08561} {arXiv:2108.08561 [hep-ph]} \BibitemShut {NoStop}%
\bibitem [{\citenamefont {Soloveva}\ \emph {et~al.}(2024)\citenamefont {Soloveva}, \citenamefont {Palermo},\ and\ \citenamefont {Bratkovskaya}}]{Soloveva:2023tvj}%
  \BibitemOpen
  \bibfield  {author} {\bibinfo {author} {\bibfnamefont {O.}~\bibnamefont {Soloveva}}, \bibinfo {author} {\bibfnamefont {A.}~\bibnamefont {Palermo}}, \ and\ \bibinfo {author} {\bibfnamefont {E.}~\bibnamefont {Bratkovskaya}},\ }\href {\doibase 10.1103/PhysRevC.110.034908} {\bibfield  {journal} {\bibinfo  {journal} {Phys. Rev. C}\ }\textbf {\bibinfo {volume} {110}},\ \bibinfo {pages} {034908} (\bibinfo {year} {2024})},\ \Eprint {http://arxiv.org/abs/2311.15984} {arXiv:2311.15984 [hep-ph]} \BibitemShut {NoStop}%
\bibitem [{\citenamefont {Li}\ \emph {et~al.}(2023)\citenamefont {Li}, \citenamefont {L\"u}, \citenamefont {Pang},\ and\ \citenamefont {Qin}}]{Li:2022ozl}%
  \BibitemOpen
  \bibfield  {author} {\bibinfo {author} {\bibfnamefont {F.-P.}\ \bibnamefont {Li}}, \bibinfo {author} {\bibfnamefont {H.-L.}\ \bibnamefont {L\"u}}, \bibinfo {author} {\bibfnamefont {L.-G.}\ \bibnamefont {Pang}}, \ and\ \bibinfo {author} {\bibfnamefont {G.-Y.}\ \bibnamefont {Qin}},\ }\href {\doibase 10.1016/j.physletb.2023.138088} {\bibfield  {journal} {\bibinfo  {journal} {Phys. Lett. B}\ }\textbf {\bibinfo {volume} {844}},\ \bibinfo {pages} {138088} (\bibinfo {year} {2023})},\ \Eprint {http://arxiv.org/abs/2211.07994} {arXiv:2211.07994 [hep-ph]} \BibitemShut {NoStop}%
\bibitem [{\citenamefont {Paszke}\ \emph {et~al.}(2019)\citenamefont {Paszke}, \citenamefont {Gross}, \citenamefont {Massa}, \citenamefont {Lerer}, \citenamefont {Bradbury}, \citenamefont {Chanan}, \citenamefont {Killeen}, \citenamefont {Lin}, \citenamefont {Gimelshein}, \citenamefont {Antiga}, \citenamefont {Desmaison}, \citenamefont {Kopf}, \citenamefont {Yang}, \citenamefont {DeVito}, \citenamefont {Raison}, \citenamefont {Tejani}, \citenamefont {Chilamkurthy}, \citenamefont {Steiner}, \citenamefont {Fang}, \citenamefont {Bai},\ and\ \citenamefont {Chintala}}]{NEURIPS2019_bdbca288}%
  \BibitemOpen
  \bibfield  {author} {\bibinfo {author} {\bibfnamefont {A.}~\bibnamefont {Paszke}}, \bibinfo {author} {\bibfnamefont {S.}~\bibnamefont {Gross}}, \bibinfo {author} {\bibfnamefont {F.}~\bibnamefont {Massa}}, \bibinfo {author} {\bibfnamefont {A.}~\bibnamefont {Lerer}}, \bibinfo {author} {\bibfnamefont {J.}~\bibnamefont {Bradbury}}, \bibinfo {author} {\bibfnamefont {G.}~\bibnamefont {Chanan}}, \bibinfo {author} {\bibfnamefont {T.}~\bibnamefont {Killeen}}, \bibinfo {author} {\bibfnamefont {Z.}~\bibnamefont {Lin}}, \bibinfo {author} {\bibfnamefont {N.}~\bibnamefont {Gimelshein}}, \bibinfo {author} {\bibfnamefont {L.}~\bibnamefont {Antiga}}, \bibinfo {author} {\bibfnamefont {A.}~\bibnamefont {Desmaison}}, \bibinfo {author} {\bibfnamefont {A.}~\bibnamefont {Kopf}}, \bibinfo {author} {\bibfnamefont {E.}~\bibnamefont {Yang}}, \bibinfo {author} {\bibfnamefont {Z.}~\bibnamefont {DeVito}}, \bibinfo {author} {\bibfnamefont {M.}~\bibnamefont {Raison}}, \bibinfo {author} {\bibfnamefont {A.}~\bibnamefont {Tejani}}, \bibinfo
  {author} {\bibfnamefont {S.}~\bibnamefont {Chilamkurthy}}, \bibinfo {author} {\bibfnamefont {B.}~\bibnamefont {Steiner}}, \bibinfo {author} {\bibfnamefont {L.}~\bibnamefont {Fang}}, \bibinfo {author} {\bibfnamefont {J.}~\bibnamefont {Bai}}, \ and\ \bibinfo {author} {\bibfnamefont {S.}~\bibnamefont {Chintala}},\ }in\ \href {https://proceedings.neurips.cc/paper_files/paper/2019/file/bdbca288fee7f92f2bfa9f7012727740-Paper.pdf} {\emph {\bibinfo {booktitle} {Advances in Neural Information Processing Systems}}},\ Vol.~\bibinfo {volume} {32},\ \bibinfo {editor} {edited by\ \bibinfo {editor} {\bibfnamefont {H.}~\bibnamefont {Wallach}}, \bibinfo {editor} {\bibfnamefont {H.}~\bibnamefont {Larochelle}}, \bibinfo {editor} {\bibfnamefont {A.}~\bibnamefont {Beygelzimer}}, \bibinfo {editor} {\bibfnamefont {F.}~\bibnamefont {d\textquotesingle Alch\'{e}-Buc}}, \bibinfo {editor} {\bibfnamefont {E.}~\bibnamefont {Fox}}, \ and\ \bibinfo {editor} {\bibfnamefont {R.}~\bibnamefont {Garnett}}}\ (\bibinfo  {publisher} {Curran
  Associates, Inc.},\ \bibinfo {year} {2019})\BibitemShut {NoStop}%
\bibitem [{\citenamefont {Kapusta}\ and\ \citenamefont {Gale}(2011)}]{Kapusta:2006pm}%
  \BibitemOpen
  \bibfield  {author} {\bibinfo {author} {\bibfnamefont {J.~I.}\ \bibnamefont {Kapusta}}\ and\ \bibinfo {author} {\bibfnamefont {C.}~\bibnamefont {Gale}},\ }\href {\doibase 10.1017/CBO9780511535130} {\emph {\bibinfo {title} {{Finite-temperature field theory: Principles and applications}}}},\ Cambridge Monographs on Mathematical Physics\ (\bibinfo  {publisher} {Cambridge University Press},\ \bibinfo {year} {2011})\BibitemShut {NoStop}%
\bibitem [{\citenamefont {Haque}\ and\ \citenamefont {Mustafa}(2025)}]{Haque:2024gva}%
  \BibitemOpen
  \bibfield  {author} {\bibinfo {author} {\bibfnamefont {N.}~\bibnamefont {Haque}}\ and\ \bibinfo {author} {\bibfnamefont {M.~G.}\ \bibnamefont {Mustafa}},\ }\href {\doibase 10.1016/j.ppnp.2024.104136} {\bibfield  {journal} {\bibinfo  {journal} {Prog. Part. Nucl. Phys.}\ }\textbf {\bibinfo {volume} {140}},\ \bibinfo {pages} {104136} (\bibinfo {year} {2025})},\ \Eprint {http://arxiv.org/abs/2404.08734} {arXiv:2404.08734 [hep-ph]} \BibitemShut {NoStop}%
\bibitem [{\citenamefont {Levai}\ and\ \citenamefont {Heinz}(1998)}]{Levai:1997yx}%
  \BibitemOpen
  \bibfield  {author} {\bibinfo {author} {\bibfnamefont {P.}~\bibnamefont {Levai}}\ and\ \bibinfo {author} {\bibfnamefont {U.~W.}\ \bibnamefont {Heinz}},\ }\href {\doibase 10.1103/PhysRevC.57.1879} {\bibfield  {journal} {\bibinfo  {journal} {Phys. Rev. C}\ }\textbf {\bibinfo {volume} {57}},\ \bibinfo {pages} {1879} (\bibinfo {year} {1998})},\ \Eprint {http://arxiv.org/abs/hep-ph/9710463} {arXiv:hep-ph/9710463} \BibitemShut {NoStop}%
\bibitem [{\citenamefont {Plumari}\ \emph {et~al.}(2011)\citenamefont {Plumari}, \citenamefont {Alberico}, \citenamefont {Greco},\ and\ \citenamefont {Ratti}}]{Plumari:2011mk}%
  \BibitemOpen
  \bibfield  {author} {\bibinfo {author} {\bibfnamefont {S.}~\bibnamefont {Plumari}}, \bibinfo {author} {\bibfnamefont {W.~M.}\ \bibnamefont {Alberico}}, \bibinfo {author} {\bibfnamefont {V.}~\bibnamefont {Greco}}, \ and\ \bibinfo {author} {\bibfnamefont {C.}~\bibnamefont {Ratti}},\ }\href {\doibase 10.1103/PhysRevD.84.094004} {\bibfield  {journal} {\bibinfo  {journal} {Phys. Rev. D}\ }\textbf {\bibinfo {volume} {84}},\ \bibinfo {pages} {094004} (\bibinfo {year} {2011})},\ \Eprint {http://arxiv.org/abs/1103.5611} {arXiv:1103.5611 [hep-ph]} \BibitemShut {NoStop}%
\bibitem [{\citenamefont {Nakamura}\ and\ \citenamefont {Sakai}(2005)}]{Nakamura:2004sy}%
  \BibitemOpen
  \bibfield  {author} {\bibinfo {author} {\bibfnamefont {A.}~\bibnamefont {Nakamura}}\ and\ \bibinfo {author} {\bibfnamefont {S.}~\bibnamefont {Sakai}},\ }\href {\doibase 10.1103/PhysRevLett.94.072305} {\bibfield  {journal} {\bibinfo  {journal} {Phys. Rev. Lett.}\ }\textbf {\bibinfo {volume} {94}},\ \bibinfo {pages} {072305} (\bibinfo {year} {2005})},\ \Eprint {http://arxiv.org/abs/hep-lat/0406009} {arXiv:hep-lat/0406009} \BibitemShut {NoStop}%
\bibitem [{\citenamefont {Astrakhantsev}\ \emph {et~al.}(2017)\citenamefont {Astrakhantsev}, \citenamefont {Braguta},\ and\ \citenamefont {Kotov}}]{Astrakhantsev:2017nrs}%
  \BibitemOpen
  \bibfield  {author} {\bibinfo {author} {\bibfnamefont {N.}~\bibnamefont {Astrakhantsev}}, \bibinfo {author} {\bibfnamefont {V.}~\bibnamefont {Braguta}}, \ and\ \bibinfo {author} {\bibfnamefont {A.}~\bibnamefont {Kotov}},\ }\href {\doibase 10.1007/JHEP04(2017)101} {\bibfield  {journal} {\bibinfo  {journal} {JHEP}\ }\textbf {\bibinfo {volume} {04}},\ \bibinfo {pages} {101} (\bibinfo {year} {2017})},\ \Eprint {http://arxiv.org/abs/1701.02266} {arXiv:1701.02266 [hep-lat]} \BibitemShut {NoStop}%
\bibitem [{\citenamefont {Meyer}(2007)}]{Meyer:2007ic}%
  \BibitemOpen
  \bibfield  {author} {\bibinfo {author} {\bibfnamefont {H.~B.}\ \bibnamefont {Meyer}},\ }\href {\doibase 10.1103/PhysRevD.76.101701} {\bibfield  {journal} {\bibinfo  {journal} {Phys. Rev. D}\ }\textbf {\bibinfo {volume} {76}},\ \bibinfo {pages} {101701} (\bibinfo {year} {2007})},\ \Eprint {http://arxiv.org/abs/0704.1801} {arXiv:0704.1801 [hep-lat]} \BibitemShut {NoStop}%
\bibitem [{\citenamefont {Andronic}\ \emph {et~al.}(2010)\citenamefont {Andronic}, \citenamefont {Braun-Munzinger},\ and\ \citenamefont {Stachel}}]{Andronic:2009jd}%
  \BibitemOpen
  \bibfield  {author} {\bibinfo {author} {\bibfnamefont {A.}~\bibnamefont {Andronic}}, \bibinfo {author} {\bibfnamefont {P.}~\bibnamefont {Braun-Munzinger}}, \ and\ \bibinfo {author} {\bibfnamefont {J.}~\bibnamefont {Stachel}},\ }\href {\doibase 10.1016/j.nuclphysa.2009.12.048} {\bibfield  {journal} {\bibinfo  {journal} {Nucl. Phys. A}\ }\textbf {\bibinfo {volume} {834}},\ \bibinfo {pages} {237C} (\bibinfo {year} {2010})},\ \Eprint {http://arxiv.org/abs/0911.4931} {arXiv:0911.4931 [nucl-th]} \BibitemShut {NoStop}%
\bibitem [{\citenamefont {Chen}\ \emph {et~al.}(2007)\citenamefont {Chen}, \citenamefont {Li}, \citenamefont {Liu},\ and\ \citenamefont {Nakano}}]{Chen:2007xe}%
  \BibitemOpen
  \bibfield  {author} {\bibinfo {author} {\bibfnamefont {J.-W.}\ \bibnamefont {Chen}}, \bibinfo {author} {\bibfnamefont {Y.-H.}\ \bibnamefont {Li}}, \bibinfo {author} {\bibfnamefont {Y.-F.}\ \bibnamefont {Liu}}, \ and\ \bibinfo {author} {\bibfnamefont {E.}~\bibnamefont {Nakano}},\ }\href {\doibase 10.1103/PhysRevD.76.114011} {\bibfield  {journal} {\bibinfo  {journal} {Phys. Rev. D}\ }\textbf {\bibinfo {volume} {76}},\ \bibinfo {pages} {114011} (\bibinfo {year} {2007})},\ \Eprint {http://arxiv.org/abs/hep-ph/0703230} {arXiv:hep-ph/0703230} \BibitemShut {NoStop}%
\bibitem [{\citenamefont {Plumari}\ \emph {et~al.}(2012)\citenamefont {Plumari}, \citenamefont {Puglisi}, \citenamefont {Scardina},\ and\ \citenamefont {Greco}}]{Plumari:2012ep}%
  \BibitemOpen
  \bibfield  {author} {\bibinfo {author} {\bibfnamefont {S.}~\bibnamefont {Plumari}}, \bibinfo {author} {\bibfnamefont {A.}~\bibnamefont {Puglisi}}, \bibinfo {author} {\bibfnamefont {F.}~\bibnamefont {Scardina}}, \ and\ \bibinfo {author} {\bibfnamefont {V.}~\bibnamefont {Greco}},\ }\href {\doibase 10.1103/PhysRevC.86.054902} {\bibfield  {journal} {\bibinfo  {journal} {Phys. Rev. C}\ }\textbf {\bibinfo {volume} {86}},\ \bibinfo {pages} {054902} (\bibinfo {year} {2012})},\ \Eprint {http://arxiv.org/abs/1208.0481} {arXiv:1208.0481 [nucl-th]} \BibitemShut {NoStop}%
\end{thebibliography}%

\end{document}